\documentclass[a4paper,11pt]{article}
\usepackage{geometry}
\usepackage{xspace}
\usepackage{url}
\usepackage[small]{caption}
\usepackage{graphicx}
\usepackage{amsmath}
\usepackage{amsthm}
\usepackage{booktabs}
\usepackage[numbers]{natbib}

\usepackage[linesnumbered,ruled,vlined]{algorithm2e}
\RequirePackage{comment}
\usepackage{mathtools}
\RequirePackage{comment}
\usepackage{xifthen}
\usepackage{arydshln}
\usepackage[hidelinks]{hyperref}
\usepackage[capitalise]{cleveref} 
\usepackage{xcolor}
\usepackage{amssymb}
\usepackage{dsfont}
\usepackage{colortbl}
\usepackage{graphics}
\usepackage{float}
\usepackage{color}
\usepackage{xcolor}
\usepackage{array}
\usepackage{multirow}
\usepackage{enumitem}

\newcommand{\CC}{\ensuremath{\mathcal C}\xspace}

\newtheorem{proposition}{\bf Proposition}

\newtheorem{theorem}{\bf Theorem}

\newtheorem{example}{\bf Example}
\newtheorem{definition}{\bf Definition}

\crefname{theorem}{theorem}{\bf Theorem}
\crefname{example}{example}{\bf Example}
\crefname{observation}{observation}{\bf Observation}
\crefname{lemma}{lemma}{\bf Lemma}
\crefname{corollary}{corollary}{\bf Corollary}
\crefname{proposition}{proposition}{\bf Proposition}
\crefname{definition}{definition}{\bf Definition}
\crefname{claim}{claim}{\bf Claim}
\crefname{reductionrule}{reduction rule}{\bf Reduction rule}

\DeclareMathOperator*{\argmaxaa}{arg\,max}
\DeclareMathOperator*{\argminaa}{arg\,min}
\newcommand{\argmi}{\argminaa\limits_}
\newcommand{\argma}{\argmaxaa\limits_}
\newcommand{\instance}{\ensuremath{I}\xspace}
\newcommand{\suml}{\sum\limits_}

\newcommand{\minl}{\min\limits_}

\DeclarePairedDelimiter\ceil{\lceil}{\rceil}

\newcommand{\curly}[1]{\ensuremath{\{#1\}}\xspace}
\newcommand{\NPH}{\ensuremath{\mathsf{NP}}-hard\xspace}

\newcommand{\WWH}{\ensuremath{\mathsf{W[2]}}-hard\xspace}

\newcommand{\subsum}{{\sc Subset Sum}\xspace}
\newcommand{\vercov}{{\sc Vertex Cover}\xspace}

\newcommand{\ccmw}{{\sc $\beta$-CC}\xspace}
\newcommand{\yes}{{\sc Yes}\xspace}
\newcommand{\no}{{\sc No}\xspace}

\newcommand{\bud}{\ensuremath{L}\xspace}
\newcommand{\proj}{\ensuremath{A}\xspace}
\newcommand{\voters}{\ensuremath{N}\xspace}
\newcommand{\feasible}{\ensuremath{\mathcal{F}}\xspace}
\newcommand{\comporder}{\ensuremath{\mathcal{L}(\proj)}\xspace}
\newcommand{\prof}{\ensuremath{\mathcal{P}}\xspace}
\newcommand{\approf}{\ensuremath{\mathcal{A}}\xspace}
\newcommand{\fullinstance}{\ensuremath{\langle \voters,\proj,c,\bud,\prof \rangle}\xspace}
\newcommand{\ii}{\ensuremath{i}\xspace}
\newcommand{\jj}{\ensuremath{j}\xspace}
\newcommand{\pa}{\ensuremath{a}\xspace}
\newcommand{\suci}{\ensuremath{\succeq_i}\xspace}

\newcommand{\cof}[1]{\ensuremath{\operatorname{c\!}\left(#1\right)}\xspace}
\newcommand{\rof}[1]{\ensuremath{\operatorname{r_i\!}\left(#1\right)}\xspace}
\newcommand{\pat}[1]{\ensuremath{\operatorname{p_i\!}\left(#1\right)}\xspace}
\newcommand{\ptill}[1]{\ensuremath{E^i_{[#1]}}\xspace}
\newcommand{\trunk}[1]{\ensuremath{\operatorname{t_i\!}\left(#1\right)}\xspace}
\newcommand{\trunkk}[2]{\ensuremath{\operatorname{t_i\!}\left(#1,#2\right)}\xspace}

\newcommand{\trunks}{\ensuremath{t}\xspace}
\newcommand{\btpara}{\ensuremath{\mu}\xspace}
\newcommand{\btr}{\ensuremath{B_\btpara}\xspace}
\newcommand{\gtr}{\ensuremath{G_f}\xspace}
\newcommand{\cwpara}{\ensuremath{\alpha}\xspace}
\newcommand{\cwr}{\ensuremath{W_{f,\cwpara}}\xspace}
\newcommand{\ebpara}{\ensuremath{\lambda}\xspace}
\newcommand{\ebr}{\ensuremath{D_\ebpara}\xspace}

\newcommand{\cwparaof}[1]{\ensuremath{\operatorname{\cwpara\!}\left(#1\right)}\xspace}
\newcommand{\bool}{\ensuremath{\mathds{1}}}
\newcommand{\br}{\ensuremath{\mathcal{R}}\xspace}
\newcommand{\winnerfunction}{\ensuremath{\Phi}\xspace}

\newcommand{\ruleof}[2]{%
  \ifthenelse{\isempty{#2}}%
    {\ensuremath{\operatorname{#1\!}\left(\instance\right)}\xspace}
    {\ensuremath{\operatorname{#1\!}\left(\instance#2\right)}\xspace}
}
\newcommand{\winners}[2]{%
  \ifthenelse{\isempty{#2}}%
    {\ensuremath{\operatorname{\winnerfunction\!}\left(#1,\instance\right)}\xspace}
    {\ensuremath{\operatorname{\winnerfunction\!}\left(#1,\instance#2\right)}\xspace}
}
\newcommand{\scoreof}[1]{\ensuremath{score(#1)}\xspace}
\newcommand{\disutilof}[1]{\ensuremath{d(#1)}\xspace}

\def\mathunderline#1#2{\color{#1}\underline{{\color{black}#2}}\color{black}}
\makeatletter
\newcommand\mathcircled[1]{%
  \mathpalette\@mathcircled{#1}%
}
\newcommand\@mathcircled[2]{%
  \tikz[baseline=(math.base)] \node[draw,circle,inner sep=1pt] (math) {$\m@th#1#2$};%
}
\makeatother
\makeatletter
\newcommand{\thickhline}{%
    \noalign {\ifnum 0=`}\fi \hrule height 1.3pt
    \futurelet \reserved@a \@xhline
}
\newcommand{\toothickhline}{%
    \noalign {\ifnum 0=`}\fi \hrule height 2pt
    \futurelet \reserved@a \@xhline
}
\newcolumntype{"}{@{\hskip\tabcolsep\vrule width 1.3pt\hskip\tabcolsep}}
\newcolumntype{x}{@{\hskip\tabcolsep\vrule width 2pt\hskip\tabcolsep}}
\makeatother

\makeatletter
\newcommand*{\rom}[1]{\expandafter\@slowromancap\romannumeral #1@}
\makeatother

\setlist[enumerate]{nosep}

\title{Indivisible Participatory Budgeting under Weak Rankings}
\author{Gogulapati Sreedurga \and Yadati Narahari}
\date{Indian Institute of Science}
\begin{document}

\maketitle

\begin{abstract}
    Participatory budgeting (PB) has attracted much attention in recent times due to its wide applicability in social choice settings. In this paper, we consider indivisible PB which involves allocating an available, limited budget to a set of indivisible projects, each having a certain cost, based on the preferences of agents over projects. The specific, important, research gap that we address in this paper is to propose classes of rules for indivisible PB with weak rankings (i.e., weak ordinal preferences) and investigate their key algorithmic and axiomatic issues. We propose two classes of rules having distinct significance and motivation. The first is \textit{layered approval rules} which enable weak rankings to be studied by carefully translating them into approval votes. The second is \textit{need-based rules} which enable to capture fairness issues. Under layered approval rules, we  study two natural families of rules: greedy-truncation rules and cost-worthy rules. The paper has two parts. In the first part, we investigate algorithmic and complexity related issues for the proposed rules. In the second part, we present a detailed axiomatic analysis of these rules, for which, we examine and generalize axioms in the literature and also introduce a new axiom, pro-affordability. The paper helps to highlight the trade-offs among practical appeal, computational complexity, and axiomatic compliance of these rules.
\end{abstract}
\section{Introduction}\label{sec: intro}
Participatory Budgeting (PB) is a popular voting method for allocating a divisible resource (time, money, etc.), called \textit{budget}, to multiple projects (also called \textit{alternatives}) by aggregating the preferences of agents over projects. Its most natural application is in democratic decision making, where a governing body allocates funds to proposed projects by aggregating the preferences of stakeholders or citizens. It has been successfully implemented in numerous social choice contexts \cite{avritzer2000public,cabannes2004participatory,rocke2014framing}.

PB is classified broadly into two categories: divisible and indivisible \cite{aziz2021participatory}. In divisible PB, the projects can be implemented fractionally and no specific costs are associated to them. Indivisible PB involves projects that are atomic and thus can only be implemented wholly. Each project has a certain associated cost  and needs to be funded fully, if selected for funding. Our work is concerned with indivisible PB.

The most studied preference elicitation method for PB is {\em approval votes\/}, in which each agent approves a subset of projects by simply announcing the subset. Divisible PB with approval votes has been widely studied under the name of {\em mixing\/} or {\em sharing\/} \cite{bogomolnaia2005collective,duddy2015fair,aziz2019fair}. Also, there is a vast body of literature that has looked into indivisible PB with approval votes \cite{goel2015knapsack,aziz2018proportionally,talmon2019framework,goel2019knapsack,jain2020participatory,jain2020participatoryg,rey2020designing,freeman2021truthful}.  Approval votes, however, have limited expressibility since the agents do not get to express their preferences among approved projects and all the approved projects are assumed to be equally desirable. Another popular preference elicitation method is strict rankings (aka strict ordinal preferences), in which each agent reports a strict ranking over the projects. Strict rankings are studied, to some extent, in divisible PB \cite{airiau2019portioning} as well as in indivisible PB \cite{shapiro2017participatory}. However, strict rankings have the limitation that they are cognitively hard for the agents to specify, especially when the number of projects is high.

\noindent {\bf PB with Weak Rankings}. In PB settings, agents usually are clear about relative utilities for the projects, based on which they have an inherent ranking of projects. This inherent ranking is usually weak, since ties among the projects are very natural, especially when the number of projects is high. This immediately motivates a study of weak rankings (aka weak ordinal preferences) in the context of PB. Weak rankings are well studied in the context of divisible PB \cite{aziz2014generalization,aziz2018rank}. However, their study in indivisible PB is rather limited. An example in this context is the paper by Aziz and Lee \cite{aziz2021proportionally} who introduced proportionality axioms in the setting of indivisible PB under weak rankings. 
Rather surprisingly, there is, as yet, no known class of PB rules with weak rankings that has been studied. We address this remarkable research gap in our work. It is notable that weak rankings are very general and subsume strict rankings and approval votes (see for example, Figure 1 in \cite{aziz2021proportionally}). This establishes the wide scope of our work. \\

\noindent {\bf Our Contributions and Outline}. As already stated, the focus of our paper is to study indivisible PB with weak rankings. We propose and study three major classes of PB rules: (1) Greedy-truncation rules; (2) Cost-worthy rules; and (3) Need-based rules. Of these, the first two classes, greedy-truncation rules and cost-worthy rules, use what we call a {\em layer\/}, that reduces  weak rankings to approval votes. Accordingly, we call them {\em layered approval rules\/}. All the three classes have a clear motivation and practical appeal which are brought out in the relevant sections.
The idea in greedy-truncation rules is to capture, for each agent, all the projects present in her most desired feasible outcomes. This can be viewed as an extension of knapsack voting \cite{goel2019knapsack} for weak rankings.
In cost-worthy rules, the idea is to capture whether or not a project is worth its cost based on the \textit{degree of preference} agents have for it. Need-based rules enable the key issue of fairness to be captured in PB. 

The paper is organized into two parts. In the first part (\Cref{sec: layeredrules} and \Cref{sec: needrules}), we motivate and introduce each of the families of rules, followed by a study of their computational complexity. In the second part (\Cref{sec: axioms}), we present a detailed axiomatic analysis of these families of rules and introduce a new, natural axiom pro-affordability. Tables \ref{tab: resultsla} and \ref{tab: resultsnb} summarise the computational and axiomatic results for all the proposed rules. These tables provide a bird's eye view of the spectrum of results derived in this paper. In our considered view, these results help fill an important gap in the study of indivisible PB with weak rankings.

\begin{table*}[t]
    \resizebox{\columnwidth}{!}{
    \begin{tabular}{c"c|c"c|c"c|c}
    Utility function \small{\textit{f(S)}} & \multicolumn{2}{c"}{$|S|$} & \multicolumn{2}{c"}{$\cof{S}$} & \multicolumn{2}{c}{$\bool(|S|>0)$}\\
    \hline
    RULES $\rightarrow$ & \gtr & \cwr & \gtr & \cwr & \gtr & \cwr\\
     \hline
     \hline
     \multirow{1}{*}{Computational} & \multirow{2}{*}{\hyperref[the: btcardinality-p]{P}} & \multirow{2}{*}{\hyperref[the: cwcardinality-p]{P}} & \multirow{2}{*}{\hyperref[the: btoverlap-nph]{\NPH}} & \multirow{1}{*}{\hyperref[the: cwoverlap-p]{poly in \cwparaof{1}}} & \multirow{2}{*}{\hyperref[the: btbool-nph]{\NPH}} & \multirow{1}{*}{\hyperref[the: cwbool-p]{P if $\cwparaof{1} = \cwparaof{m}$}}\\
      \multirow{1}{*}{Complexity} & & & & \multirow{1}{*}{\hyperref[the: cwoverlap-nph]{\NPH if $\cwparaof{1} = \bud$}} & & \multirow{1}{*}{\hyperref[the: cwbool-nph]{\NPH if $\cwparaof{1} \neq \cwparaof{m}$}}\\
     \hline
     \hline
    \hyperref[def: candidatem]{Candidate} & \multirow{2}{*}{\hyperref[the: candidatem]{$\checkmark$}} & \multirow{2}{*}{\hyperref[the: candidatem]{$\checkmark$}} & \multirow{2}{*}{\hyperref[the: candidatem]{$\checkmark$}} & \multirow{2}{*}{\hyperref[the: candidatem]{$\checkmark$}} & \multirow{2}{*}{\hyperref[the: candidatem]{$\checkmark$}} & \multirow{2}{*}{\hyperref[the: candidatem]{$\checkmark$}}\\
    \hyperref[def: candidatem]{Monotonicity} & & & & & & \\
    \hline
    \multirow{1}{*}{\hyperref[def: noncrossingm]{Non-crossing}} & \multirow{2}{*}{\hyperref[the: noncrossingm]{$\checkmark$}} & \multirow{2}{*}{\hyperref[the: noncrossingm]{$\checkmark$}} & \multirow{2}{*}{\hyperref[the: noncrossingm]{$\checkmark$}} & \multirow{2}{*}{\hyperref[the: noncrossingm]{$\checkmark$}} & \multirow{2}{*}{\hyperref[prop: noncrossingm]{$\times$}} & \multirow{1}{*}{\hyperref[the: noncrossingm]{$\checkmark$} if $\cwparaof{1} = \cwparaof{m}$}\\
     \multirow{1}{*}{\hyperref[def: noncrossingm]{Monotonicity}} & & & & & & \multirow{1}{*}{\hyperref[prop: noncrossingm]{$\times$} if $\cwparaof{1} \neq \cwparaof{m}$}\\
     \hline
    \multirow{1}{*}{\hyperref[def: discountm]{Discount}} & \multirow{2}{*}{\hyperref[prop: noncrossingm]{$\times$}} & \multirow{2}{*}{\hyperref[the: noncrossingm]{$\checkmark$}} & \multirow{2}{*}{\hyperref[prop: noncrossingm]{$\times$}} & \multirow{2}{*}{\hyperref[prop: noncrossingm]{$\times$}} & \multirow{2}{*}{\hyperref[prop: noncrossingm]{$\times$}} & \multirow{2}{*}{\hyperref[the: noncrossingm]{$\checkmark$}}\\
     \multirow{1}{*}{\hyperref[def: discountm]{Monotonicity}} & & & & & & \\
     \hline
     \multirow{1}{*}{\hyperref[def: limitm]{Limit}} & \multirow{2}{*}{\hyperref[the: limitm]{$\times$}} & \multirow{1}{*}{\hyperref[prop: limitm]{$\checkmark$} if $\cwparaof{1} < \cwparaof{m}+2$} & \multirow{2}{*}{\hyperref[the: limitm]{$\times$}} & \multirow{1}{*}{\hyperref[prop: limitm]{$\checkmark$} if $\cwparaof{1} \leq 1$} & \multirow{2}{*}{\hyperref[the: limitm]{$\times$}} & \multirow{1}{*}{\hyperref[prop: limitm]{$\checkmark$} if $\cwparaof{1} < \cwparaof{m}+2$}\\
     \multirow{1}{*}{\hyperref[def: limitm]{Monotonicity}} & & \multirow{1}{*}{\hyperref[the: limitm]{$\times$} if $\cwparaof{1} \geq \cwparaof{m}+2$} & & \multirow{1}{*}{\hyperref[the: limitm]{$\times$} if $\cwparaof{1} > 1$} & & \multirow{1}{*}{\hyperref[the: limitm]{$\times$} if $\cwparaof{1} \geq \cwparaof{m}+2$}\\
     \hline
    \multirow{1}{*}{\hyperref[def: proafford]{Pro-affordability}} & \multirow{1}{*}{\hyperref[the: proafford]{$\checkmark$}} & \multirow{1}{*}{\hyperref[the: proafford]{$\checkmark$}} & \multirow{1}{*}{\hyperref[prop: proafford]{$\times$}} & \multirow{1}{*}{\hyperref[prop: limitm]{$\checkmark$} if and only if $\cwparaof{1} \leq 1$} & \multirow{1}{*}{\hyperref[the: proafford]{$\checkmark$}} & \multirow{1}{*}{\hyperref[the: proafford]{$\checkmark$}}\\
     \hline
    \multirow{1}{*}{Others} & \multirow{1}{*}{$\checkmark$} & \multirow{1}{*}{$\checkmark$} & \multirow{1}{*}{$\checkmark$} & \multirow{1}{*}{$\checkmark$} & \multirow{1}{*}{$\checkmark$} & \multirow{1}{*}{$\checkmark$}\\
    \end{tabular}
    }
    \caption{A summary of our results for \textbf{layered approval rules}. \gtr stands for greed-truncation rule and \cwr stands for cost-worthy rule with the parameter $\cwpara$. {\em Others\/} include \textbf{five properties}: anonymity, neutrality, consistency (Section \ref{sec: genericaxioms}), \hyperref[def: splittingm]{splitting monotonicity}, and \hyperref[def: inclusionmax]{inclusion maximality} (Section \ref{subsec: pbaxioms}). Notations $c,S,$ and \bud are as defined in \Cref{sec: notations}.}
    \label{tab: resultsla}
\end{table*}
\vspace*{-\baselineskip}
\begin{table*}[t]
    \centering
    \scalebox{0.8}{
    \begin{tabular}{c"c|c|c}
    PARAMETER $\ebpara \in$ & $(0,1]$ & $(1,\bud-1]$ & $(\bud-1,\bud]$\\
     \hline
     \hline
     Computational Complexity & \multicolumn{3}{c}{\multirow{1}{*}{\hyperref[the: ebr-w2h]{\WWH}}}\\
     \hline
     \hline
    \hyperref[def: candidatem]{Candidate Montonicity} & \multirow{1}{*}{\hyperref[the: candidatem]{$\checkmark$}} & \multirow{1}{*}{\hyperref[prop: candidatem]{$\times$}} & \multirow{1}{*}{\hyperref[the: candidatem]{$\checkmark$}}\\
    \hline
    \hyperref[def: noncrossingm]{Non-crossing Monotonicity} & \multirow{1}{*}{\hyperref[prop: noncrossingm]{$\times$}} & \multirow{1}{*}{\hyperref[prop: noncrossingm]{$\times$}} & \multirow{1}{*}{\hyperref[prop: noncrossingm]{$\times$}}\\
    \hline
    \hyperref[def: discountm]{Discount Monotonicity} & \multirow{1}{*}{\hyperref[the: discountm]{$\checkmark$}} & \multirow{1}{*}{\hyperref[prop: discountm]{$\times$}} & \multirow{1}{*}{\hyperref[prop: discountm]{$\times$}}\\
    \hline
    \hyperref[def: limitm]{Limit Monotonicity} & \multirow{1}{*}{\hyperref[the: limitm]{$\times$}} & \multirow{1}{*}{\hyperref[the: limitm]{$\times$}} & \multirow{1}{*}{\hyperref[the: limitm]{$\times$}}\\
    \hline
    \hyperref[def: proafford]{Pro-affordability} & \hyperref[the: proafford]{$\checkmark$} & \hyperref[prop: proafford]{$\times$} & \hyperref[prop: proafford]{$\times$}\\
    \hline
    {Others} & \multirow{1}{*}{$\checkmark$} & \multirow{1}{*}{$\checkmark$} & \multirow{1}{*}{$\checkmark$}\\
    \end{tabular}
    }
    \caption{A summary of our results for \textbf{need-based rule} \ebr. \bud is budget.\emph{Others} include \textbf{five properties}: anonymity, neutrality, consistency (Sec.\ref{sec: genericaxioms}), \hyperref[def: splittingm]{splitting monotonicity}, and \hyperref[def: inclusionmax]{inclusion maximality} (Sec.\ref{subsec: pbaxioms}).}
    \label{tab: resultsnb}
\end{table*}
\section{Preliminaries and Notations}\label{sec: notations}
Let $\voters = \{1,\ldots,n\}$ denote the set of $n$ agents and $\proj = \{a_1,\ldots,a_m\}$ denote the set of $m$ projects. Each agent \ii has a ranking $\suci \in \comporder$ where \comporder denotes the set of all complete weak orders over \proj. That is, each \suci partitions \proj into equivalence classes such that $E^i_1 \!\succ_i\! E^i_2 \!\succ_i\! \ldots$ and the rank of a project $a$ is said to be $r$ if $a \in E^i_r$. A preference profile \prof is a vector of rankings of all agents in \voters, i.e., $\prof = {(\suci)}_{i \in \voters}$. A cost function $c: \proj \to \mathbb{Z}^+$ gives the cost of each project. The cost of a subset $S \subseteq \proj$, $\sum_{a \in S}{\cof{a}}$, is denoted by \cof{S}. Let \bud be the total budget available. Any subset $S \subseteq \proj$ is said to be \textit{feasible} if $c(S) \leq \bud$. Let \feasible be the set of all feasible subsets of projects.

For each weak ranking \suci in \prof, we define $\operatorname{r_i}$ and $\ptill{j}$ as follows: 
\begin{itemize}
    \item for some $\pa \in \proj$, ${\rof{\pa}}$ denotes the rank of project \pa in \suci, i.e., $\pa \in E^i_{\rof{\pa}}$.
    \item for some $j \in [m]$, ${\ptill{j}}$ denotes the set of projects whose rank is at most \jj in \suci, i.e., $\ptill{j} = \bigcup_{k \leq j} E^i_k$
\end{itemize}

\begin{example}
Suppose $A= \{a_1,a_2,a_3,a_4\}$ is the set of projects such that $\cof{a_1} = 3, \cof{a_2} = 4, \cof{a_3} = 2$, and $\cof{a_4} = 6$. Let the ranking of a certain agent, say $i$, be $\curly{a_2, a_3} \!\succ_i\! \curly{a_1} \!\succ_i\! \curly{a_4}$. Now, $\rof{a_3} = 1, \ptill{2} = E^i_1\! \cup\! E^i_2 = \curly{a_1,a_2,a_3}$.
\end{example}

An ordinal PB instance  \fullinstance is represented by \instance. All the rules we present in this paper are {\em irresolute\/}, which implies that the rules output one or more feasible subsets. For an instance \instance, a PB rule \br outputs a set of feasible subsets of projects, denoted by \ruleof{\br}{}.
\section{Layered Approval Rules}\label{sec: layeredrules}
We call this class {\em layered approval rules\/} to indicate that these rules reduce the weak rankings to approval votes with the help of components called {\em layers\/}. The purpose of the layers is to exploit the expressiveness of weak rankings to capture certain key information and use this information to convert weak rankings into approval votes. 
Thus, the layered approval rules harness the computational and cognitive simplicity of approval voting and may be viewed as a hybrid of ranking and approval voting.

Consider PB with approval votes. Suppose each agent $i$ approves a set of projects $A_i$. The utility of agent $i$ from a set of projects $S$ is defined as $u_i(S) = f(A_i \cap S)$, where $f$ is what we call a general utility function. There are three such popular general utility functions in the literature \cite{talmon2019framework}: $f(S) = |S|$,  $f(S) = \cof{S}$, and $f(S) = \bool(|S| > 0)$ (takes $1$ if $S$ is non-empty and $0$ otherwise). Layered approval rules with any of the above utility functions involve two stages: (i) a layer reduces the weak ranking profile into approval vote profile based on certain key information (ii) for the resultant approval vote profile, the rule computes feasible sets of projects that maximize the sum of utilities of all the agents. We explore two classes of layered approval votes, namely, greedy-truncation rules and cost-worthy rules.

\subsection{Greedy-Truncation Rules}\label{sec: gtr}
Greedy-truncation rules extend the idea behind knapsack votes (a subclass of approval votes) to PB with weak rankings. The standard approval votes do not take the budget constraint into account and thus do not capture the information about the feasible subsets considered ideal by each agent. To resolve this problem, Goel et al. \cite{goel2019knapsack} introduced knapsack voting, where for every agent $i$, $\cof{A_i} \leq \bud$. That is, every agent approves her most desired feasible subset of projects. However, knapsack voting does not perform well when there are ties among the projects and there are multiple equally desirable feasible subsets for the agent. The proposed greedy-truncation rules solve this problem by generalizing the idea behind knapsack voting to consider weak rankings, as illustrated by the following example.

\begin{example}\label{eg1: btr}
Let  $A = \{a_1,a_2,\ldots,a_{8}\}$  and $\bud = 5$, with costs $\{3,3,2,\ldots,2\}$, respectively. Let there be two agents whose inherent preferences for projects are as follows:
\vspace*{-0.8\baselineskip}
\begin{table}[H]
    \scalebox{0.9}{
    \begin{tabular}{ccccc!{\color{orange}\vrule}cccccccccc}
    $a_1$ & $\succ_1$ & \{$a_3$ & $,$ & $a_4$ & $,$ & $a_2$\} & $\succ_1$ & $a_5$ & $\succ_1$ & \{$a_6$ & $,$ & $a_7$\} & $\succ_1$ & $a_{8}$\\
    \end{tabular}}\\
    \scalebox{0.9}{
    \begin{tabular}{ccc!{\color{orange}\vrule}cccccccccccc}
    \{$a_1$ & $,$ & $a_3$\} & $\succ_2$ & \{$a_4$ & $,$ & $a_6$\} & $\succ_2$ & $a_5$ & $\succ_2$ & \{$a_2$ & $,$ & $a_7$\} & $\succ_2$ & $a_{8}$\\
    \end{tabular}}
\end{table}
\vspace*{-0.8\baselineskip}
First, suppose the agents are asked for standard approval votes and agent $1$ approves the set $A_1 = \{a_1,a_2,a_3,a_4,a_5\}$. Notice that the fact that the agent 1 prefers the feasible outcome $\{a_1,a_4\}$ over $\{a_1,a_5\}$ is lost. Now, suppose the agents are asked to report knapsack votes. Then, agent $1$ is forced to approve $a_1$ and exactly one project from $\{a_3,a_4\}$, thereby implying incorrectly that the remaining project yields no utility to her (note that both $\curly{a_1,a_3}$ and $\curly{a_1,a_4}$ are equally desirable feasible outcomes).

Greedy-truncation rules overcome this problem by allowing weak rankings and including all the projects with ties, while respecting the budget constraint. The layer truncates the preference to include exactly $\curly{a_1,a_3} \cup \curly{a_3,a_4} = \curly{a_1,a_3,a_4}$, as indicated by the red markers. \qed
\end{example}
\vspace*{-0.3\baselineskip}
The layer in greedy-truncation rules greedily adds the projects following the rank until the budget constraint is respected. At the point where the total cost crosses \bud, the layer includes only those projects whose inclusion will not make the subset infeasible (in the above example, $a_2$ is not included since $\curly{a_1,a_2}$ is infeasible). We now formally explain the layer with a pseudocode.
\vspace*{-\baselineskip}
\begin{algorithm}[h!]
\DontPrintSemicolon
\KwIn{Ranking profile \prof, budget \bud}
\KwOut{Approval vote profile \approf}
\For{each agent $i$}{
    $A_i \gets \emptyset$; $Z \gets 0$\\
    $k_i \gets \text{number of equivalence classes in }\succeq_i$\;
    \For{$j = 1 ; j\leq k_i; j++$}{
        \If{$\cof{E^i_j}+\cof{A_i} \leq \bud$}{
            $A_i \gets A_i$ $\cup E^i_j$;             $Z \gets Z+\cof{E^i_j}$\;
        }
        \Else{
            \For{\text{each }$a$\text{ in }$E^i_j$}{
                \If{$\cof{a} \leq \bud - Z$}{
                    $A_i \gets A_i$ $\cup \{a\}$\;
                }
            }
            \break\;
        }
    }
}
\Return{$(A_i)_{i \in [n]}$}\;
\caption{Greedy-truncation layer}
\label{algo:truncation}
\end{algorithm}
\vspace*{-\baselineskip}

It needs to be mentioned that adding the projects greedily w.r.t. rank may not always result in an optimal knapsack vote. However, solving the knapsack problem is \NPH and cognitively hard for the agents \cite{benade2021preference}. Thus, especially when the costs of projects are close to each other, the greedy approach provides a close-to-ideal outcome with much less cognitive load.

\begin{definition}[Greedy-Truncation Rules]\label{def: btr}
A greedy-truncation rule \gtr with the general utility function $f: 2^A \to \mathbb{R}_{\geq 0}$ outputs \vspace*{-0.7\baselineskip}$$\left\{S^*: S^* \in \argma{S \in \feasible}{\sum_{i \in \voters}{f\big({A_i\;\cap\;S}\big)}}\right\}$$
where $A_i$ is the approval vote of agent $i$ in the vote profile obtained from \Cref{algo:truncation}.
\end{definition}

\begin{example}\label{eg2: btr}
Consider Example \ref{eg1: btr}. Clearly, after executing the layer as explained in \Cref{algo:truncation}, we get $A_1 = \{a_1,a_3,a_4\}$ and $A_2 = \{a_1,a_3\}$. If $f(S)$ is defined as $|S|$ or $\cof{S}$, the  greedy-truncation rule outputs $\curly{\{a_1,a_3\}}$. If $f(S)$ is defined as $\bool(|S| > 0)$, then the  greedy-truncation rule outputs all the sets containing at least one of $a_1$ and $a_3$.
\end{example}

It is notable that the greedy-truncation rules are generalizations of existing multi-winner voting rules \cite{faliszewski2018multiwinner}.
Represent a ordinal multi-winner voting instance as a PB instance with $\bud = k$ and unit cost projects. If $f(S)$ is $|S|$ or $\cof{S}$, $G_f$ is equivalent to multi-winner bloc rule. If $f(S)\! =\! \bool(|S| > 0)$, $G_f$ is equivalent to $\alpha_k$-CC rule. If $k\! =\! 1$, then any $G_f$ reduces to single-winner plurality rule. We now analyze the computational complexity of \gtr in the following theorems.
\begin{theorem}\label{the: btoverlap-nph}
If $f(S) = c(S)$, $G_f$ is \NPH.
\end{theorem}
\begin{proof}
Let $\approf = (A_i)_{i \in [n]}$ be the approval-vote profile obtained from the weak rankings profile \prof using \Cref{algo:truncation}. We first reduce \subsum to the decision version of $G_f$, which is to decide if there exists $S \in \feasible$ such that $\sum_{i \in \voters}{f\big({A_i\;\cap\;S}\big)} \geq s$ for any score $s$. Given an integer $Z$ and a set of integers $X= \{x_1,x_2,\ldots,x_n\}$, the \subsum problem is to decide if there exists a subset $X' \subseteq X$ such that $\sum_{x \in X'}{x} = Z$. This is known to be \NPH \cite{garey1979computers}. W.l.o.g., we can assume that $X$ is sorted in non-decreasing order. Given an instance as above, we construct a PB instance as follows: set $\bud = s = Z$. Create $n$ projects $a_1,a_2,\ldots,a_n$ such that $c(a_i) = x_i$ and another project $a_{n+1}$ with $c(a_{n+1}) = Z$. Create $n$ agents such that the preference of agent \ii is $a_i \succ_i a_{n+1} \succ_i others$ (`others' could be ordered arbitrarily). We claim that both these problems are equivalent.

For the constructed PB instance, \Cref{algo:truncation} approves only the top-ranked project for each agent \ii. Hence, for each \ii, $A_i$ is $\{a_i\}$. To prove our claim, let us first assume that the given instance is a \yes instance of \subsum. Let the subset $X'$ be the required subset. For a set of projects $S = \{a_i : x_i \in X'\}$, $\sum_{i \in \voters}{f\big({A_i\;\cap\;S}\big)} = \sum_{x_i \in X'}{x_i} = Z$. Hence, this is a \yes instance of the given problem. Now, let us assume that the given instance is a \no instance of \subsum. Therefore, there is no subset $X' \subseteq X$ such that $\sum_{x_i \in X'}{x_i} = Z$. For any $S \in \feasible$, $\sum_{i \in \voters}{f\big({A_i\;\cap\;S}\big)} < s$, making this a \no instance of the given problem. Since the decision version of \gtr is proved to be \NPH, it follows that \gtr is \NPH (else, decision version could have been solved in polynomial time).
\end{proof}

\begin{theorem}\label{the: btbool-nph}
If $f(S) = \bool(|S| > 0)$, $G_f$ is \NPH.
\end{theorem}
\begin{proof}
Let $\approf = (A_i)_{i \in [n]}$ be the approval-vote profile obtained from the weak rankings profile \prof using \Cref{algo:truncation}. We first reduce \vercov to the decision version of \btr, which is to decide if there exists $S \in \feasible$ such that $\sum_{i \in \voters}{f\big({A_i\;\cap\;S}\big)} \geq s$ for any score $s$. Given an undirected graph $G = (V,E)$ and an integer $k$, the \vercov problem is to decide if there exists a $V' \subseteq V$ such that $|V'| \leq k$ and $E = \{(v_1,v_2): (v_1 \in V') \lor (v_2 \in V')\}$. We construct a PB instance as follows. Set $\bud = k$. For each vertex $v$, add a project $a_v$ with cost $1$. Add another dummy project $a_d$ with cost $k-1$. For each edge $e_i = (v^1_i,v^2_i)$, add an agent $i$ with preference $a_{v^1_i} \succ_i a_{v^2_i} \succ_i a_d \succ_i others$ (`others' could be ordered arbitrarily). Set $s = |E|$. We claim that both these instances are equivalent.

We can assume w.l.o.g. that $k > 2$ (for constant values of $k$, the \vercov becomes tractable). Therefore for each agent $i$, \Cref{algo:truncation} gives an output such that $A_i = \{a_{v^1_i},a_{v^2_i}\}$. To prove our claim, assume that the given instance $(G,k)$ is a \yes instance of \vercov. Let $V' \subseteq V$ cover entire $E$ such that $|V'| \leq k$. Consider the feasible subset of projects $S = \{a_v: v \in V'\}$. Since $V'$ is a vertex cover, for any agent \ii, $|A_i \cap S| > 0$. Therefore, $\sum_{i \in \voters}{f\big({A_i\;\cap\;S}\big)} = |E|$. Hence, this is a \yes instance. Now, let us assume that $(G,k)$ is a \no instance. Any feasible subset of projects gives a total score less than $|E|$ and the instance is thus a \no instance. Since the decision version of \gtr is proved to be \NPH, it follows that \gtr is \NPH (else, decision version could have been solved in polynomial time).
\end{proof}

\begin{theorem}\label{the: btcardinality-p}
If $f(S) = |S|$, $G_f$ is polynomial-time computable.
\end{theorem}
\begin{proof}
Let $\approf = (A_i)_{i \in [n]}$ be the approval-vote profile obtained from the weak rankings profile \prof using \Cref{algo:truncation}. This result is directly implied by an existing result for participatory budgeting with approval votes \cite{talmon2019framework}. Here, we present the dynamic programming based algorithm.

Let us define score of a project $a$ as $score(a) = |\{\;i \in \voters: a \in A_i\;\}|$. Construct a table $T$ with $m$ rows and $mn$ columns. The entry $T(i,j)$ corresponds to the cost of cheapest subset of $\{a_1,\ldots,a_i\}$ for which the total score of projects is exactly $j$. We fill the first row as follows:
\[T(1,j) = \begin{cases} 
      c(a_1) & j = score(a_1) \\
      0 & \text{otherwise} 
   \end{cases} \quad \forall j \in [mn].
\]
Now the remaining rows can be filled recursively as follows:
\begin{equation}
    \label{eqn: btcardinality-recurrence}
    T(i,j) = \min\Big\{T\big(i-1,j\big)\;,\;T\big(i-1,j-score(a_i)\big)+c(a_i)\Big\}
\end{equation}
The set of projects corresponding to each $T(i,j)$ can be stored in another matrix by doing the following: if the former value is minimum in \cref{eqn: btcardinality-recurrence}, $a_j$ will not be added to the set; otherwise, it will be added. The maximum score possible is $\max\{j \in [mn]: T(m,j) \leq \bud\}$ and the corresponding set of projects is the outcome of \btr. Note that computing each entry of $T$ takes constant time and there are $m^2n$ entries. The running time is $O(m^2n)$. Correctness follows from the definition of $T$.
\end{proof}

\subsection{Cost-Worthy Rules}
The  layer in these rules captures whether or not a project is worth its cost, based on the \textit{degree of preference} the agents have for the project.

In many applications, there is a natural preference for projects that are economical. Some PB rules for approval votes treat all the approved projects the same \cite{jain2020participatory,jain2020participatoryg,rey2020designing} while others prefer the expensive ones assuming that the cost reflects its prestige and quality  \cite{goel2015knapsack,aziz2018proportionally,talmon2019framework,goel2019knapsack,freeman2021truthful}. We note that even the greedy-truncation rules we studied in \Cref{sec: gtr} do not prefer an inexpensive project over an expensive one, when both are equally preferred by all agents. While this could be reasonable in some contexts, often there are real-world scenarios in which inexpensive projects are to be preferred and an expensive project is to be funded only if it is preferred over other projects by a high enough number of agents. This requirement cannot be captured in approval votes since the preference among the approved projects remains unknown, whereas it can be captured in weak rankings. The following is a motivating example.

\begin{example}\label{motieg: cwr}
Consider a funding agency that wishes to fund a few research projects using a budget $100$. An expert panel of 6 experts evaluates the project applications based on technical novelty, impact, etc. Three of the experts submit the same ranking of projects $A\!\succ \! B \!\succ\! \{C , D\} \!\succ\! E$ while the other three experts submit the same ranking $A\! \succ \!C \!\succ\! \{B, D\} \!\succ\! E$. Suppose the costs of $\{A,B,C,D,E\}$ are $\{10,90,30,30,30\}$ respectively. A greedy-truncation rule will output $\{A,B\}$ or $\{A,C\}$ or $\curly{\{A,B\},\{A,C\}}$. While $B$ and $C$ are ranked similarly by the experts, $\cof{C}$ is only $1/3$rd of $\cof{B}$. So, pragmatically, $\{A,C,D,E\}$ would have been a better choice.
\end{example}

Our layer in cost-worthy rules can be used to capture this requirement in weak rankings. One can argue that it is better to directly ask an agent to approve only the projects she thinks are worth their costs, instead of using our layer. However if we do so, the agent will simply compare the expected output of the project with the cost. Whereas, with the proposed layer, a planner can decide the worth of a project by considering the \textit{degree} of preference of agents for the project. For example, say a certain agent considers a prohibitively expensive project to be worthy of its cost but yet less preferable over three other projects. Our layer offers the planner the flexibility to decide that the fourth rank is not a good enough rank to warrant such a huge cost.

We use a parameter $\cwpara \in \{0,1,\ldots,\bud\}^m$ such that $\cwpara(i) \geq \cwpara(j)$ whenever $i < j$. We call this parameter \emph{worth vector} and use this to capture the relation between ranks and the costs of the projects. For any $j \in [m]$, $\cwpara(j)$ denotes the maximum amount a project ranked at position $j$ deserves. From an agent's perspective, $\cwpara$ indicates whether or not she ranks each project high enough to justify its cost. The following example explains this parameter.

\begin{example}\label{eg: cwr}
Suppose the organizers of a two hour seminar must select a subset of 8 candidate talks (each of different duration) based on the preferences of audience. They may decide that only one of the talks, the plenary talk, be given a duration of 60 min., while the other talks be allocated at most 40 or 20 min. Based on this, they could set the parameter as $\cwpara = (60,40,40,40,20,20,20,0)$. Alternatively, they could  formulate the duration deserved by a talk as a function of its rank.
\end{example}

The layer approves, for each agent, only those projects that are affordable according to their ranks and the worth vector. This is explained formally in \Cref{algo:worthy}.
\vspace*{-0.8\baselineskip}
\begin{algorithm}[h!]
\DontPrintSemicolon
\KwIn{Ranking profile \prof, parameter \cwpara}
\KwOut{Approval vote profile \approf}
\For{each agent $i$}{
    $A_i \gets \emptyset$\\
    \For{$j = 1 ; j\leq m; j++$}{
        \If{$\cof{a_j} \leq \cwpara(r_i(a_j))$}{
            $A_i \gets A_i$ $\cup \{a_j\}$\;
        }
    }
}
\Return{$(A_i)_{i \in [n]}$}\;
\caption{Cost-worthy layer}
\label{algo:worthy}
\end{algorithm}
\vspace*{-\baselineskip}
\begin{definition}[Cost-worthy rules]\label{def: cwr}
A cost-worthy rule \cwr with the general utility function $f: 2^A \!\to\! \mathbb{R}_{\geq 0}$ and a parameter $\cwpara\! \in\! \{0,1,\ldots,\bud\}^m$ such that $\cwpara(i) \!\geq\! \cwpara(j)$ for every $i < j$, outputs \vspace*{-0.7\baselineskip}$$\left\{S^*: S^* \in \argma{S \in \feasible}{\sum_{i \in \voters}{f\big({A_i\;\cap\;S}\big)}}\right\}$$
where $A_i$ is the approval vote of agent $i$ in the vote profile obtained from \Cref{algo:worthy}.
\end{definition}

In other words, a cost-worthy rule maximizes the sum of scores of the projects, where the score of a project is the number of agents who rank it high enough.

\begin{example}
Consider the instance described in \cref{motieg: cwr} and a cost-worthy rule \cwr with $\cwpara = (100,80,60,60,60,0,\ldots,0)$.
\begin{table}[H]
    \scalebox{0.85}{
    \begin{tabular}{c|cccccc}
    $\boldsymbol{\alpha}$ & $\!\!\mathbf{100}\:\;\;\;$ & $\!\!\mathbf{80}\:\;\;\;$ & $\!\!\mathbf{60}\:\;\;\;$ & $\!\!\mathbf{60}\:\;\;\;$ & $\mathbf{0}$\\
    \hline
    $50\%$ & $\textcolor{orange}{A}\:\;\;\succ$ & $B\:\;\;\succ$ & $\{\textcolor{orange}{C},\textcolor{orange}{D}\}\:\;\;\succ$ & $\textcolor{orange}{E}\:\;\;\succ$ & $\ldots$\\
    {\small{of experts}}& $\!\!10\:\;\;\;$ & $\!\!90\:\;\;\;$ & $\!\!30,30\:\;\;\;$ & $\!\!30\:\;\;\;$ & $\ldots$\\
    \hline
    $50\%$ & $\textcolor{orange}{A}\:\;\;\succ$ & $\textcolor{orange}{C}\:\;\;\succ$ & $\{B,\textcolor{orange}{D}\}\:\;\;\succ$ & $\textcolor{orange}{E}\:\;\;\succ$ & $\ldots$\\
    {\small{of experts}}& $\!\!10\:\;\;\;$ & $\!\!30\:\;\;\;$ & $\!\!90,30\:\;\;\;$ & $\!\!30\:\;\;\;$ & $\ldots$\\
    \end{tabular}
    }
\end{table}
\vspace*{-\baselineskip}
All projects ranked satisfactorily high enough are marked above. $B$ is not ranked high enough to justify its cost. If $f(S)$ is defined as $|S|$ or $\cof{S}$, the corresponding cost-worthy rule outputs $\{A,C,D,E\}$. If $f(S)$ is defined as $\bool(|S| > 0)$, the corresponding cost-worthy rule outputs all the sets containing at least one of $A$, $C$, $D$, and $E$.
\end{example}

We describe a few situations where cost-worthy rules provide a desirable choice, by specifying the corresponding worth vectors. Consider an instance where a funding agency wants no project to be allocated an amount higher than $x$ unless it is of the highest quality. This can be captured using $(\bud,x,\ldots,x)$. If the funding agency believes that the projects in the top $p$ positions have stiff competition among themselves and there is a consensus regarding the remaining projects, a function whose top $p$ entries are almost similar while the rest of the entries have larger gaps can be the parameter. If the projects ranked less than a certain threshold should never be approved, we can achieve this by setting to zero, all the entries in the worth vector after the threshold. The situations above show multiple ways in which this family of rules provides flexibility to the social planner. In \Cref{sec: axioms}, we establish that this family of rules satisfies almost all the desirable axioms.

It is notable that while the parameters for cost-worthy rules can be decided intuitively and heuristically in the real-world, there is an interesting optimization problem underlying their choice. Learning optimal parameters of desirable rules such as median and min-max rules and rules with threshold approval votes \cite{benade2021preference} are considered important open directions in the voting literature. Similarly, the question of determining optimal parameters for cost-worthy rules is an important problem which we preserve for our future work. Next, we analyze the computational complexity of \cwr.

\begin{proposition}\label{the: cwoverlap-p}
If $f(S) = c(S)$, the complexity of solving \cwr is polynomial in $\cwpara(1)$.
\end{proposition}
\begin{proof}
Let $\approf = (A_i)_{i \in [n]}$ be the approval-vote profile obtained from the weak rankings profile \prof using \Cref{algo:worthy}.

We present a dynamic programming algorithm. Let us define score of a project $a$ as $score(a) = |\{\;i \in \voters: a \in A_i\;\}|*\cof{a}$. Construct a table $T$ with $m$ rows and $mn\cwpara(1)$ columns. The entry $T(i,j)$ corresponds to the cost of a cheapest subset of $\{a_1,\ldots,a_i\}$ for which the total score of projects is exactly $j$. $T$ can be constructed using the same procedure as explained in the proof of \Cref{the: btcardinality-p}. Note that whenever $a \in A_i$ for some $i$, it follows that $\cof{a} \leq \cwpara(r_i(a)) \leq \cwpara(1)$.

Note that computing each entry of $T$ takes constant time and there are $m^2n\cwpara(1)$ entries. The running time is $O(m^2n\cwpara(1))$. Correctness follows from the definition of $T$.
\end{proof}

The above proposition implies that when $f(S) = c(S)$, for small or constant values of $\cwpara(1)$, \cwr is polynomial-time computable. Below, we show that when $\cwpara(1)$ is as large as \bud, \cwr becomes computationally hard.

\begin{theorem}\label{the: cwoverlap-nph}
If $f(S) = c(S)$, \cwr is \NPH for any worth vector \cwpara with $\cwpara(1) = \bud$.
\end{theorem}
\begin{proof}
Let $\approf = (A_i)_{i \in [n]}$ be the approval-vote profile obtained from the weak rankings profile \prof using \Cref{algo:worthy}. We again reduce \subsum to the decision version of \cwr as in \Cref{the: btoverlap-nph}.

Given a \subsum instance, we construct a PB instance as follows. Set $\bud = s = Z$. Create $n$ projects $a_1,a_2,\ldots,a_n$ such that $c(a_i) = x_i$. Create a single agent who ranks all these $n$ projects in the first place. We claim that both these problems are equivalent.

For the constructed PB instance, \Cref{algo:worthy} approves all the projects $a_1,a_2,\ldots,a_n$. Rest of the proof follows from the proof of \Cref{the: btoverlap-nph}.

Since the decision version of \cwr is proved to be \NPH, it follows that \cwr is \NPH (else, decision version could have been solved in polynomial time).
\end{proof}

\begin{proposition}\label{the: cwbool-p}
If $f(S) = \bool(|S| > 0)$, \cwr is polynomial-time computable for any worth vector \cwpara with $\cwpara(1) = \cwpara(m)$.
\end{proposition}
\begin{proof}
Suppose $\cwpara(1) = \cwpara(m) = Q$. That is, in the output of \Cref{algo:worthy}, all the agents approve all and only those projects whose cost is at most $Q$. Hence any such project, when it exists, will be selected by \cwr as an optimal solution.
\end{proof}

\begin{theorem}\label{the: cwbool-nph}
If $f(S) = \bool(|S| > 0)$, \cwr is \NPH for any worth vector \cwpara with $\cwpara(1) \neq \cwpara(m)$.
\end{theorem}
\begin{proof}
Let $\approf = (A_i)_{i \in [n]}$ be the approval-vote profile obtained from the weak rankings profile \prof using \Cref{algo:worthy}.

We again reduce \vercov to the decision version of \cwr by constructing a PB instance as follows: set $\bud = k(\cwpara(m)+1)$. For each vertex $v$, add a project $a_v$ with cost $\cwpara(m)+1$. Add $m-2$ dummy projects $d_1,\ldots,d_{m-2}$ such that $\cof{d_i} = \cwpara(i+1) + 1$. For each edge $e_i = (v^1_i,v^2_i)$, add an agent $i$ with preference $\{a_{v^1_i}, a_{v^2_i}\} \succ_i d_1 \succ_i \ldots \succ_i d_{m-2} \succ_i others$. Set $s = |E|$. We claim that both these instances are equivalent.

We can assume w.l.o.g. that $k > 2$ (for constant values of $k$, the \vercov becomes tractable). Since $\cwpara(1) \geq \cwpara(m)+1$, for each agent $i$, \Cref{algo:worthy} gives an output such that $A_i = \{a_{v^1_i},a_{v^2_i}\}$. Rest of the proof follows from the proof of \Cref{the: btbool-nph}.

Since the decision version of \cwr is proved to be \NPH, it follows that \cwr is \NPH (else, decision version could have been solved in polynomial time).
\end{proof}

The proof of the next result proceeds on the same lines of \Cref{the: btcardinality-p}, except that \Cref{algo:worthy} is used to compute \approf in the place of \Cref{algo:truncation}.
\begin{proposition}\label{the: cwcardinality-p}
If $f(S) = |S|$, \cwr is polynomial-time computable for any worth vector \cwpara.
\end{proposition}

\section{Need-Based Rules}\label{sec: needrules}
Need-based rules are suitable for capturing several facets of {\em fairness\/}.  While the existing fairness notions in indivisible PB literature mostly deal with proportionality \cite{aziz2018proportionally,peters2020proportional,aziz2021proportionally,pierczynski2021proportional,fairstein2021proportional}, fairness in divisible PB is based on guarantees for each agent at individual level (or in other words, \emph{need-based}). In many real-world scenarios, an agent will be happy if a  certain minimum fraction of the budget is spent on projects favorable to her. This is captured in divisible PB by \emph{fair share} \cite{bogomolnaia2005collective,aziz2019fair,aziz2014generalization,aziz2018rank,airiau2019portioning}. The notion of individual fair share \cite{bogomolnaia2005collective}, for example, ensures that for every agent, at least ${\frac{1}{n}}\scriptstyle{th}$ of the budget is allocated to projects favored by the agent.

Need-based rules capture such a requirement for indivisible PB. The parameter $\ebpara \in (0,\bud]$, called \emph{need}, denotes the amount needed to make an agent happy. A need-based rule with parameter \ebpara is denoted by \ebr. The happiness of an agent depends on how \ebpara is allocated. For the need \ebpara and a set $S \subseteq \proj$, let us define \trunkk{\ebpara}{S} as follows:
    \vspace{-0.5em}
    { \begin{align*}
        \trunkk{\ebpara}{S} \!=\! \begin{cases} 
      \min \big\{j \!\in\! [m]: \cof{\ptill{j} \cap S} \!\geq \!\ebpara\big\}\!\! & \!\text{if }\cof{S}\! \!\geq\!\! \ebpara \\
      m+1 & \text{otherwise} 
   \end{cases}
    \end{align*}}
A solution where at least \ebpara is allocated to projects in \ptill{\trunkk{\ebpara}{A}} is ideal for agent $i$. Instead, if a set $S$ is chosen, the amount \ebpara is not allocated till $E^i_{[\trunkk{\ebpara}{S}]}$. The disutility of agent $i$ from a set $S$ is hence $\trunkk{\ebpara}{S} - \trunkk{\ebpara}{A}$. Our proposed rule \ebr minimizes the total disutility of the agents. Since \trunkk{\ebpara}{A} is independent of $S$, it can be safely dropped from the optimization objective.

\begin{definition}[Need-based rules]\label{def: ebr}
A need-based rule \ebr with a parameter $\ebpara \in (0,\bud]$ outputs \vspace*{-0.3\baselineskip}$$\left\{S^*: S^* \in \argmi{S \in \feasible}{\suml{i \in \voters}{\trunkk{\ebpara}{S}}}\right\}.$$\vspace*{-\baselineskip}
\end{definition}

Intuitively appealing need parameters include $\ebpara = \frac{\bud}{n}$ and $\ebpara = \bud$. The former reflects the idea of individual fair share \cite{bogomolnaia2005collective} in indivisible PB whereas the latter reflects the situation where an agent is unhappy if even one unit of the budget is spent on a project ranked very low.

\begin{example}
Let  $A = \{a_1,a_2,a_3,a_4,a_5\}$ and $\bud = 12$, with costs $4,2,5,3,$ and $2$ respectively. Let us consider a need-based rule \ebr with $\ebpara = 7$. Suppose there are two agents whose preferences are as follows:
\vspace*{-\baselineskip}
\begin{table}[H]
    \delimitershortfall=0pt
    \setlength{\dashlinegap}{1pt}
    \scalebox{0.9}{
    \begin{tabular}{ccc!{\color{orange}\vrule}cc!{\color{blue}\vrule}cc}
    $\mathunderline{blue}{\textcolor{orange}{a_1}}$ & $\succ_1$ & $\{\mathunderline{blue}{a_2},\textcolor{orange}{a_4}\}$ & $\succ_1$ & $\mathunderline{blue}{\textcolor{orange}{a_3}}$ & $\succ_1$ & $a_5$\\
    $4$ & & $2,\;3$ & & $5$ & & $2$\\
    \end{tabular}}\\
    ~\\
    \scalebox{0.9}{
    \begin{tabular}{c!{\color{orange}\vrule}cc!{\color{blue}\vrule}cccc}
    $\{\mathunderline{blue}{\textcolor{orange}{a_3}},\textcolor{orange}{a_4}\}$ & $\succ_2$ & $\mathunderline{blue}{\textcolor{orange}{a_1}}$ & $\succ_2$ & $a_5$ & $\succ_2$ & $\mathunderline{blue}{a_2}$\\
    $5,\;3$ & & $4$ & & $2$ & & $2$\\
    \end{tabular}}
\end{table}
\vspace*{-\baselineskip}
\noindent{Consider two feasible sets $S_1 = \curly{a_1,a_2,a_3}$ and $S_2 = \curly{a_1,a_3,a_4}$. Projects in $S_1$ are underlined in blue and projects in $S_2$ are colored in orange. For agent $2$ and set $S_1$, the disutility $\trunks_2(7,S_1) = 2$ since $\cof{a_3} < 7$ but $\cof{a_1}+\cof{a_3} > 7$. The points at which need is allocated are marked for both agents and sets. Clearly, $S_1$ has a total disutility of $3+2 = 5$ while $S_2$ has $2+1=3$. In fact, $S_2$ has the optimal disutility and is output by \ebr.}
\end{example}

\begin{theorem}\label{the: ebr-w2h}
Deciding if there exists $S \in \feasible$ such that $\sum_{i \in \voters}{\trunkk{\ebpara}{S}} \leq s$ for a given score $s$ is \WWH when parameterized by \ebpara.
\end{theorem}
\begin{proof}
We give a parameterized reduction from \ccmw multi-winner voting rule to \ebr. Given a set \CC of candidates, a set of voters \voters with their ranking profile \prof, an integer $k$, and a value $s$, the problem of \ccmw is to decide if there exists a $C' \subseteq C$ of size at most $k$ such that $\suml{i \in \voters}{(\minl{c \in C'}{\rof{c}})} \leq s$. This is known to be \WWH when parameterized by $k$ \cite{betzler2013computation}.

Given an instance of \ccmw, we construct a PB instance as follows: for each $c_i \in \CC$, create a project $a_i$ with $c(a_i) = 1$. Let the set of voters, their ranking profile, and the value $s$ remain the same. Set $\bud = k$. Set the parameter of \ebr rule as $\ebpara = 1$. Note that this is a polynomial time parameterized reduction since $k \geq 1$ by definition. We claim that both these instances are equivalent.

Note that since the cost of any project is $1$, for any agent $i$ and a set $S \in \feasible$, $\trunkk{\ebpara}{S} = \min_{a \in S}{\rof{a}}$. To prove our claim, first assume that the given \ccmw instance is a \yes instance. This implies, $\exists C' \subseteq C$ of size at most $k$ such that $\suml{i \in \voters}{(\minl{c \in C'}{\rof{c}})} \leq s$. Consider $S = \{a_i: c_i \in C'\}$. Note that $S \in \feasible$ since $c(S) = k = \bud$.
Therefore, $\sum_{i \in \voters}{\trunkk{\ebpara}{S}} \leq s$ and this is a \yes instance. Now, assume that given \ccmw instance is a \no instance. This implies $\forall C' \subseteq C$ of size at most $k$, $\suml{i \in \voters}{(\minl{c \in C'}{\rof{c}})} > s$. A subset $C' \subseteq C$ is of size at most $k$ if and only if $S = \{a_i: c_i \in C'\}$ is feasible. Therefore for any $S \in \feasible$, $\sum_{i \in \voters}{\trunkk{\ebpara}{S}} > s$ and this is a \no instance. This completes the proof.
\end{proof}

\section{Axiomatic Analysis of Proposed Rules}\label{sec: axioms}
Here, we undertake an axiomatic analysis to gain additional insights and conclude the section with some crucial observations. Throughout this section, for any given PB rule \br and an instance \instance, we use \winners{\br}{} to denote the set of all projects included in the output of \br. Formally, $\winners{\br}{} = \{a \in \proj: \exists S \in \ruleof{\br}{} a \in S\}$. We say a project $a$ wins if and only if $a \in \winners{\br}{}$. For any layered approval rule \br with the general utility function $f(S) = |S|$, we use $\scoreof{a}$ to denote the score of a project $a\!\! \in\!\! \proj$, i.e., $\scoreof{a} = |i: a \in A_i|$. If $f(S) = c(S)$, $\scoreof{a}$ is $\cof{a}*|i: a \in A_i|$. For rules with any of the above general utility functions, the utility of a set $S$ of projects is said to be the sum of scores of all projects in the set. If $f(S) = \bool(|S|>0)$, the utility of a set $S$ is equal to $|i :S \cap A_i \neq \emptyset|$. For the rule \ebr, the disutility of a set $S$ is equal to $\sum_{i \in \voters}{\trunks_i(\ebpara,S)}$.  Note that layered approval rules select a set of projects with maximum utility while need-based rules select a set with minimum disutility.

\subsection{Axioms for a Generic Voting Model}\label{sec: genericaxioms}
First, we examine the axioms in the literature applicable for any voting model \cite{brandt2016handbook,skowron2019axiomatic,lackner2021consistent}. The first two axioms we study ensure an impartial treatment of all the agents and alternatives. They require that the outcome of a rule should not depend on the indexing of agents or alternatives. Let $\Sigma_S$ be the set of all permutations on a set $S$.

\begin{definition}[Anonymity]\label{def: anonymity}
A PB rule \br is said to be anonymous if for any instance \instance and $\sigma \in \Sigma_\voters$, we have $\br(\instance)=\br(\instance')$ whenever $\instance'$ is obtained from \instance by replacing $\succ_i$ with $\succ_{\sigma(i)}$ for every $i \in \voters$.
\end{definition}
\begin{definition}[Neutrality]\label{def: neutrality}
A PB rule \br is said to be neutral if for any instance \instance and $\sigma \in \Sigma_\proj$, we have $\br(\instance)=\br(\instance')$ whenever $\instance'$ is obtained from \instance by replacing \cof{j} with \cof{\sigma(j)} and \pat{j} with \pat{\sigma(j)} in every $\succ_i$ for every $j \in [m]$.
\end{definition}
\begin{proposition}\label{prop: anonymityneutrality}
All greedy-truncation rules, cost-worthy rules, and need-based rules are anonymous and neutral.
\end{proposition}
Anonymity follows since all the rules are utilitarian and permuting the agents does not affect the summation. Neutrality follows since all rules depend only on the cost of projects and their ranks in preferences. The next popularly studied axiom, consistency, requires that for a social choice rule, if two disjoint groups of agents $N_1$ and $N_2$ both choose an outcome $a$, $a$ continues to be chosen if the groups participate together in an election, i.e., for $N_1 \cup N_2$.

\begin{definition}[Consistency]\label{def: consistency}
A PB rule \br is said to be consistent if for any two ordinal PB instances $\instance_1 = \langle \voters_1,\proj,c,\bud,\prof_1 \rangle$ and $\instance_2 =\langle \voters_2,\proj,c,\bud,\prof_2 \rangle$ such that $\voters_1 \cap \voters_2 = \emptyset$, we have, $$\br(\instance_1) \cap \br(\instance_2) \subseteq \br(\langle \voters_1 \cup \voters_2,\proj,c,\bud,\prof_1 \cup \prof_2 \rangle).$$
\end{definition}
\begin{proposition}\label{prop: consistency}
All greedy-truncation rules, cost-worthy rules, and need-based rules are consistent.
\end{proposition}
\begin{proof}
Consider any $\instance_1$ and $\instance_2$ as given in \cref{def: consistency}. Take any greedy-truncation rule \gtr, a set $S \in \gtr(\instance_1) \cap \gtr(\instance_2)$, and the instance $\instance' = \langle \voters_1 \cup \voters_2,\proj,c,\bud,\prof_1 \cup \prof_2 \rangle$. Since $S \in \gtr(\instance_1)$, $\sum_{i \in \voters_1}{f\big({A_i\;\cap\;S}\big)} \geq \sum_{i \in \voters_1}{f\big({A_i\;\cap\;S'}\big)}$ for any $S' \subseteq \proj$. Likewise, $\sum_{i \in \voters_2}{f\big({A_i\;\cap\;S}\big)} \geq \sum_{i \in \voters_2}{f\big({A_i\;\cap\;S'}\big)}$. 
By including both these inequalities, we get $S \in \ruleof{\gtr}{'}$.
Proofs for cost-worthy rules and need-based rules follow the same idea.
\end{proof}

\subsection{Axioms for Strict Rankings}\label{subsec: mwordinalaxioms}
Here, we examine the axioms proposed for voting with strict rankings. These are studied for multi-winner voting and can be extended to any voting model with strict rankings. We choose only the axioms which also make sense in the context of PB and see if our rules defined for weak rankings satisfy them. The first axiom, candidate monotonicity, ensures that if a winning project is shifted forward in any preference by one position, the project continues to win.

\begin{definition}[Candidate Monotonicity]\label{def: candidatem}
A PB rule \br satisfies candidate monotonicity if for any instance \instance and a project $x \in \winners{\br}{}$, we have $x \in \winners{\br}{'}$ whenever $\instance'$ is obtained from \instance by shifting $x$ one position forward in some preference $\succ_i$ \textup{\cite{elkind2017properties}}.
\end{definition}
\begin{theorem}\label{the: candidatem}
All the following rules satisfy candidate monotonicity:
\begin{enumerate}
    \itemsep0em
    \item Any greedy-truncation rule or cost-worthy rule where $f(S)$ is $|S|$, $\cof{S}$, or $\bool(|S|>0)$
    \item Any need-based rule \ebr where $\ebpara \in (0,1] \cup (\bud-1,\bud]$
\end{enumerate}
\end{theorem}
\begin{proof}
For any instance \instance and a PB rule \br, let $x$ be a project such that $x \in \winners{\br}{}$. Consider any $\succ_i \in \prof$. Let $j = \rof{x}$ and $x' = \pat{j-1}$. Construct $\instance'$ by exchanging $x$ and $x'$ in $\succ_i$.

We start by considering layered approval rules. First, let $f(S)$ be $|S|$ or $\cof{S}$. It can be observed that in $\instance'$, \scoreof{x} will increase or remain the same, whereas, \scoreof{x'} will remain the same or decrease. This is because, for greedy-truncation rules, depending on \cof{x} and \cof{x'}, the point of truncating the preference will either stay the same in $\instance'$ or change from $j$ to $j-1$ or from $j-1$ to $j$. For cost-worthy rules, depending on \cwparaof{j}, \cwparaof{j-1}, \cof{x}, and \cof{x'}, $x$ will be approved by the same agents and possibly by one more agent. Likewise, $x$ will be approved by the same number of agents or possibly by exactly one agent less than that number. The scores of all other projects will remain the same in $\instance'$. Since $x \in \winners{\br}{}$, there exists $S_x \in \ruleof{\br}{}$ such that $x \in S_x$. By definition, the utility of any set is the sum of scores of all projects in that set. Hence, some set containing $x$ will continue to win. Therefore, $x \in \winners{\br}{'}$. 

Now, say $f(S) = \bool(|S|>0)$. The utility of any set $S$ is the number of agents who have some project of $S \in A_i$. The utility of any $S_x \in \ruleof{\br}{}$ increases by $1$ in $\instance'$ (when $x$ is the only project in $S_x$ in $A_i$) or remains the same, otherwise. Similarly, the utility of any set without $x$ stays the same or decreases by $1$ in $\instance'$. Therefore, $x \in \winners{\br}{'}$.

Finally, we look at need-based rules. Suppose \br is a need-based rule \ebr such that $\ebpara \in (0,1]$. For any set $S$ and agent $l$, $\trunks_l(\ebpara,S)$ is the rank of best ranked project of $S$ in $\succ_l$. Disutility of $S$, \disutilof{S}, can change only because of \trunkk{\ebpara}{S} since preferences of all the other agents are unperturbed. Consider any set $S$. Let $x \in S$ and $x' \notin S$. If $\trunkk{\ebpara}{S} = j$ in \instance, then $\trunkk{\ebpara}{S} = j-1$ in $\instance'$. Else if $\trunkk{\ebpara}{S} < j$ in \instance, then $\trunkk{\ebpara}{S}$ remains the same in $\instance'$. Note that in \instance, $\trunkk{\ebpara}{S} \leq j$ always as $x \in S$. Hence, $\disutilof{S}$ in $\instance'$ is at most that in $\instance$. Now, say $x \notin S$ and $x' \in S$. If $\trunkk{\ebpara}{S} = j-1$ in \instance, then $\trunkk{\ebpara}{S} = j$ in $\instance'$. Else if $\trunkk{\ebpara}{S} < j-1$ in \instance, then $\trunkk{\ebpara}{S}$ remains the same in $\instance'$. Note that in \instance, $\trunkk{\ebpara}{S} \leq j-1$ always as $x' \in S$. Hence, $\disutilof{S}$ in $\instance'$ is at least that in $\instance$. If both $x$ and $x'$ are present or absent together in $S$, $\trunkk{\ebpara}{S}$ and \disutilof{S} are always the same in both \instance and $\instance'$. Therefore, disutility of any set with $x$ in $\instance'$ will be at most that in \instance and disutility of remaining sets will remain the same or increase. Since $x \in \winners{\ebr}{}$, some set with $x$ continues to win and $x \in \winners{\ebr}{'}$. The case of $\ebpara \in (\bud-1,\bud]$ can be argued similarly.
\end{proof}
\begin{proposition}\label{prop: candidatem}
No need-based rule \ebr such that $\ebpara \in (1,\bud-1]$ satisfies candidate monotonicity.
\end{proposition}
\begin{proof}
Consider \ebr such that $\ebpara \in (1,\bud-1]$. Consider a PB instance \instance with budget $\bud$ and $\proj = \curly{a_1, a_2, a_3}$ costing $\bud, \bud-1,$ and $1$ respectively. Let there be two agents whose preferences are as follows:
\begin{table}[H]
    \centering
    \scalebox{0.8}{
    \begin{tabular}{c|c}
         \multirow{5}{*}{\instance}&    \begin{tabular}{c!{\color{orange}\vrule}cc!{\color{blue}\vrule}cc}
        $a_1$ & $\succ_1$ & $a_2$ & $\succ_1$ & $a_3$\\
        $\scriptstyle{\bud}$ & & $\scriptscriptstyle{\bud-1}$ & & $1$\\
        \end{tabular}\\
         & \begin{tabular}{c!{\color{blue}\vrule}cc!{\color{orange}\vrule}cc}
        $a_2$ & $\succ_2$ & $a_1$ & $\succ_2$ & $a_3$\\
       $\scriptscriptstyle{\bud-1}$ & & $\scriptstyle{\bud}$ & & $1$\\
    \end{tabular}\\~\\
        \hline
        ~\\
         \multirow{4}{*}{$\instance'$}&    \begin{tabular}{c!{\color{orange}\vrule}cccc}
        $a_1$ & $\succ_1$ & $a_3$ & $\succ_1$ & $a_2$\\
        $\scriptstyle{\bud}$ & & $1$ & & $\scriptscriptstyle{\bud-1}$\\
        \end{tabular}\\
         & \begin{tabular}{ccc!{\color{orange}\vrule}cc}
        $a_2$ & $\succ_2$ & $a_1$ & $\succ_2$ & $a_3$\\
       $\scriptscriptstyle{\bud-1}$ & & $\scriptstyle{\bud}$ & & $1$\\
    \end{tabular}\\
    \end{tabular}
    }
\end{table}
Since $\ebpara \in (1,\bud-1]$, $\ruleof{\ebr}{} = \curly{\curly{a_1}, \curly{a_2}, \curly{a_2,a_3}}$. The points at which the need is allocated to the agents are marked in orange for \curly{a_1} and blue for the other two sets. Thus, $a_3 \in \winners{\ebr}{}$. $a_2$ and $a_3$ are interchanged in $\succ_1$ to get $\instance'$. Clearly, $\ruleof{\ebr}{'} = \curly{\curly{a_1}}$ and $a_3 \notin \winners{\ebr}{'}$.
\end{proof}

The next axiom, non-crossing monotonicity, ensures that if some project in a winning set is shifted forward without disturbing the  other projects in the set, the set continues to win.

\begin{definition}[Non-crossing Monotonicity]\label{def: noncrossingm}
A PB rule \br satisfies non-crossing monotonicity if for any instance \instance, a set $S \in \ruleof{\br}{}$, and a project $x \in S$, we have $S \in \ruleof{\br}{'}$ whenever $\instance'$ is obtained from \instance by shifting $x$ one position forward in some preference $\succ_i$ in which $x$ is ranked immediately below $x' \notin S$   \textup{\cite{elkind2017properties}}.
\end{definition}
\begin{proposition}\label{the: noncrossingm}
Any greedy-truncation rule or cost-worthy rule with $f(S)$ as $|S|$ or $\cof{S}$ satisfies non-crossing monotonicity. Also, the cost-worthy rule \cwr with $f(S) = \bool(|S|>0)$ satisfies non-crossing monotonicity when $\cwpara(1) = \cwpara(m)$.
\end{proposition}
\begin{proof}
For any instance \instance and a PB rule \br, let $S_x$ be a set such that $S_x \in \ruleof{\br}{}$ and $x \in S_x$. Let $\succ_i \in \prof$ be a preference such that $E^i_{\rof{x}-1} \notin S_x$. Let $j = \rof{x}$ and $x' = E^i_{j-1}$. Construct $\instance'$ by exchanging $x$ and $x'$ in $\succ_i$. For the sake of contradiction, assume $S_x \notin \ruleof{\br}{'}$.

The proof is an extension to the proof of \Cref{the: candidatem} and we use the fact that for all the said rules, the utility of a set can be expressed as the sum of scores of projects in the set. We give an outline of the argument. From \Cref{the: candidatem}, we know that the utility of $S_x$ increases or stays the same. We also know that there exists $S \in \ruleof{\br}{'}$ such that $x \in S$. This is possible only if \scoreof{x} increases since $S_x \notin \ruleof{\br}{'}$. But if \scoreof{x} increases, since $x' \notin S_x$, utility of $S_x$ also increases. This implies $S_x \in \ruleof{\br}{'}$. Now, consider the rule \cwr when $\cwparaof{1} = \cwparaof{m}$. Let $\cwparaof{1} = Q$. All agents approve all and only the projects whose cost is at most $Q$. Hence, the outcome is independent of the ranks of the projects in the preferences and the result follows.
\end{proof}

\begin{theorem}\phantomsection\label{prop: noncrossingm}
The following rules do not satisfy non-crossing monotonicity:
\begin{enumerate}
    \itemsep0em
    \item The greedy-truncation rule or any cost-worthy rule \cwr with $\cwparaof{1} \neq \cwparaof{m}$ when $f(S)$ is $\bool(|S| > 0)$.
    \item Any need-based rule \ebr
\end{enumerate}
\end{theorem}
\begin{proof}
First, let \br be the greedy-truncation rule \gtr. Consider an instance \instance with budget \bud$(>2)$, a set of projects $\proj = \curly{a_1,a_2,a_3,a_4,a_5,a_6}$ costing $1$, $\bud-1$, $\bud-1$, $\bud$, $1$, and $\bud$ respectively, and three agents whose preferences are: $a_1\!\! \succ_1\!\! a_2 \!\!\succ_1\!\! a_3 \!\!\succ_1 \!\!others$, $a_3\!\! \succ_2\!\! a_6\!\! \succ_2 \!\!others$, and $a_5\!\! \succ_3\!\! a_4 \succ_3 others$ respectively. Clearly, $\ruleof{\br}{} = \curly{\curly{a_1,a_3},\curly{a_2,a_5},\curly{a_3,a_5},\curly{a_1,a_5}}$. Now let $S = \curly{a_1,a_3}$. See that $a_2 \notin S$. Exchange $a_2$ and $a_3$ in $\succ_1$ to obtain a new instance $\instance'$. In $\instance'$, $\curly{a_3,a_5}$ has a strictly higher utility than $S$ and hence $S \notin \ruleof{\br}{'}$.

Now, let \br be the cost-worthy rule \cwr with $\cwparaof{1} \neq \cwparaof{m}$. This implies that there exists some $t \in \{1,\ldots,m\}$ such that $\cwparaof{t} > \cwparaof{t+1}$. Consider an instance \instance with budget $\bud = \cwparaof{1}+\cwparaof{t}$ and a set of projects $\proj = \curly{a_1,\ldots,a_5,d_1,\ldots,d_{m-5}}$ whose costs are as follows: $\cof{a_1} = \cof{a_4} = \cwparaof{1}$; $\cof{a_2} = \cof{a_3} = \cwparaof{t}$; $\cof{a_5} = \cwparaof{1}+\cwparaof{t}$; for every $i \in \curly{1,\ldots,t-2}$ $\cof{d_i} = \cwparaof{i+1}+1$; for every $i  \in \curly{t-1,\ldots,m-5}$, $\cof{d_i} = \cwparaof{i+3}+1$. Suppose we have three agents whose preferences are (1) $a_1 \succ d_1 \succ \ldots \succ d_{t-2} \succ a_2 \succ a_3 \succ d_{t-1} \succ \ldots \succ d_{m-5} \succ a_5 \succ a_4$ (2) $a_4 \succ d_1 \succ \ldots \succ d_{t-2} \succ a_2 \succ a_3 \succ d_{t-1} \succ \ldots \succ d_{m-5} \succ a_5 \succ a_1$ and (3) $a_5 \succ d_1 \succ \ldots \succ d_{t-2} \succ a_3 \succ a_2 \succ d_{t-1} \succ \ldots \succ d_{m-5} \succ a_4 \succ a_1$. Since $\cwparaof{t} > \cwparaof{t+1}$, \Cref{algo:worthy} gives $A_1 = \curly{a_1,a_2}$, $A_2 = \curly{a_4,a_2}$, $A_3$ is either $\curly{a_5,a_3}$ or $\curly{a_3}$. Hence, the outcome of \cwr with $f(S) = \bool(|S|>0)$ is $\curly{\curly{a_1,a_2},\curly{a_1,a_3},\curly{a_4,a_2},\curly{a_4,a_3}}$ since each of these sets has an utility of exactly $2$. Now, let $S = \curly{a_1,a_3}$. See that $a_2 \notin S$. Exchange $a_2$ and $a_3$ in the first preference to obtain new instance $\instance'$. In $\instance'$, $\curly{a_3,a_4}$ has a strictly better utility than $S$ and hence $S \notin \ruleof{\cwr}{'}$.

Now, consider a need-based rule \ebr. First we prove that if $\ebpara$ is independent of \bud and is a constant, it is easy to construct an instance where \ebr does not satisfy non-crossing monotonicity. To see this\footnote{This example is similar to that given in Prop. 9 by Elkind et al. \cite{elkind2017properties} for $l_1$-CC}, take an instance \instance with $\bud = 2 \ceil{\ebpara}$, 10 projects $\curly{a,b,c,d,x_1,\ldots,x_6}$ each costing $\ceil{\ebpara}$, six agents with the following preferences: (i) $a \succ x_1 \succ c \succ b \succ d \succ others$ (ii) $a \succ x_2 \succ d \succ b \succ c \succ others$ (iii) $b \succ x_3 \succ a \succ c \succ d \succ others$ (iv) $b \succ x_4 \succ d \succ c \succ a \succ others$ (v) $c \succ x_5 \succ a \succ b \succ d \succ others$ and (vi) $c \succ x_6 \succ d \succ b \succ a \succ others$. Clearly, $\ruleof{\ebr}{} = \curly{\curly{a,b},\curly{a,c},\curly{b,c}}$. Now let $S = \curly{a,c}$. Shift $c$ one position forward in first preference to get $\instance'$. Here, disutility of $\curly{b,c}$ is strictly less than that of \curly{a,c} and hence $\curly{a,c} \notin \ruleof{\ebr}{'}$.

Now, we construct three examples, respectively, to prove that \ebr does not satisfy non-crossing monotonicity when $\ebpara$ is in $(0,\frac{\bud}{2}]$, $(\frac{\bud}{2},\bud-1)$, and $[\bud-1,\bud]$. First, assume $\ebpara \leq \frac{\bud}{2}$. From the above paragraph, we can assume without loss of generality that $\ebpara > 2$. Consider an instance with budget \bud, five projects $\curly{a,b,x_1,x_2,x_3}$ costing $\bud-\ebpara$, $2$, $\ebpara-1$, $\bud+2$, and $\bud+2$, and two agents whose preferences are $a \succ_1 b \succ_1 x_2 \succ_1 x_1 \succ_1 x_3$ and $x_1 \succ_2 b \succ_2 x_2 \succ_2 x_3 \succ_2 a$, respectively. Since $\ebpara \leq \frac{\bud}{2}$, $\bud-\ebpara \geq \ebpara$. Therefore, $\curly{a,x_1}$ and $\curly{b,x_1}$ are present in \ruleof{\ebr}{} as each of them has optimal disutility of $6$. After interchanging $x_1$ and $x_2$ in $\succ_1$, only $\curly{b,x_1}$ wins and $\curly{a,x_1}$ does not win anymore. This completes the argument.

Now, assume $\ebpara \in (\frac{\bud}{2},\bud-1)$. Consider an instance with budget \bud, four projects $\curly{a,b,x_1,x_2}$ costing $\bud\!-\!\ebpara$, $\bud\!-\!\ebpara$, $\ebpara\!-\!1$, and $\bud\!+\!2$, and two agents whose preferences are $b \succ_1 x_2 \succ_1 x_1 \succ_1 a$ and $x_1 \succ_2 a \succ_2 b \succ_2 x_2$ respectively. Since $\ebpara < \bud-1$, $\curly{a,x_1,b} \notin \feasible$. And since $\ebpara > \frac{\bud}{2}$, $\bud-\ebpara < \ebpara$. Therefore, $\ruleof{\ebr}{} = \curly{\curly{a,x_1},\curly{b,x_1}}$ with each set having optimal disutility of $6$. On exchanging $x_1$ and $x_2$ in $\succ_1$ to get $\instance'$, $\ruleof{\ebr}{'} \!=\! \curly{\curly{b,x_1}}$. So, $\curly{a,x_1}$ does not win anymore.

Finally, assume $\ebpara > \bud-1$. Consider an instance with budget \bud, five projects $\curly{a,b,x_1,x_2,x_3}$ costing $\bud-4$, $\bud-2$, $2$, $\bud+2$, and $2$, and two agents whose preferences are $a \succ_1 b \succ_1 x_2 \succ_1 x_1 \succ_1 x_3$ and $x_1 \succ_2 a \succ_2 x_3 \succ_2 b \succ_2 x_2$, respectively. Since $\ebpara > \bud-1$, $\ruleof{\ebr}{} = \curly{\curly{a,x_1,x_3},\curly{b,x_1}}$. If we exchange $x_1$ and $x_2$ in $\succ_1$ to get $\instance'$, $\ruleof{\ebr}{'} \!=\! \curly{\curly{b,x_1}}$. That is, $\curly{a,x_1,x_3}\notin \winners{\ebr}{'}$.
\end{proof}

\subsection{Axioms for Indivisible PB}\label{subsec: pbaxioms}
In this section, we extend indivisible PB axioms defined for approval votes by Talmon and Faliszewski \cite{talmon2019framework} to weak rankings. We also introduce a new axiom, pro-affordability, for indivisible PB under weak rankings. The first axiom, splitting monotonicity, requires that, if a winning project is replaced by a set of multiple new projects that together cost the same, then at least one of the new projects continues to win.

\begin{definition}[Splitting Monotonicity]\label{def: splittingm}
A PB rule \br satisfies splitting monotonicity if for any instance \instance and $x \in \winners{\br}{}$, we have $X \cap \winners{\br}{'} \neq \emptyset$ whenever $\instance'$ is obtained from $\instance$ by spliting $x$ into a set $X$ of new projects such that $\cof{x} = \cof{X}$ and replacing $x$ by $X$ in every $\succeq_i$ in \prof.
\end{definition}
\begin{theorem}\label{the: splittingm}
All the following rules satisfy splitting monotonicity:
\begin{enumerate}
    \itemsep0em
    \item Any greedy-truncation rule or cost-worthy rule where $f(S)$ is $|S|$, $\cof{S}$, or $\bool(|S|>0)$
    \item Any need-based rule \ebr
\end{enumerate}
\end{theorem}
\begin{proof}
For an instance \instance and PB rule \br, let $x$ be a project in \winners{\br}{} and let $S_x$ be a set such that $S_x \in \ruleof{\br}{}$ and $x \in S_x$. Let $X$ be a new set of projects such that $\cof{X} = \cof{x}$. Let $\instance'$ be an instance obtained by replacing $x$ with $X$ in every ranking $\suci \in \prof$. Let $S' = (S_x \setminus \curly{x}) \cup X$. It is enough to prove that $S' \in \ruleof{\br}{'}$.

First, we look at the layered approval rules. We prove that for any $x_1 \in X$, if $x \in A_i$ in \instance, then $x_1 \in A_i$ in $\instance'$. This follows for the greedy-truncation rules since the truncation point remains the same in both the instances. Since $\cof{x_1} < \cof{x}$ and $x \in A_i$ in \instance, $x_1 \in A_i$ in $\instance'$. This also follows for the cost-worthy rules since $\cwparaof{\rof{x}}$ in \instance is same as $\cwparaof{\rof{x_1}}$ in $\instance'$. Since $\cof{x_1} < \cof{x}$ and $x \in A_i$ in \instance, $x_1 \in A_i$ in $\instance'$.

Suppose \br is \gtr or \cwr with $f(S)$ being $|S|$ or $\cof{S}$. For any $x_1 \in X$, we know from the above that \scoreof{x} = \scoreof{x_1}. This implies, if $f(S) = |S|$, total utility of $S'$ in $\instance'$ is exactly $|X|-1$ times the total utility of $S_x$ in $\instance$. If $f(S) = \cof{S}$, for any $x_1 \in X$, we know from the above that \scoreof{x_1} = (\cof{x_1}/\cof{x}) \scoreof{x}. Therefore, total utility of $S'$ in $\instance'$ is the same as the total utility of $S_x$ in $\instance$. Now, say $f(S) = \bool(|S|>0)$. From the above, the number of agents who have some project of $S_x$ in $A_i$ is the same as those who have some project of $S'$ before it. In all the above scenarios, since $S_x \in \ruleof{\br}{}$, $S' \cap \ruleof{\br}{'} \neq \emptyset$.

Finally, let \br be a need-based rule \ebr. For any $\ebpara \in (0, \bud]$ and agent \ii, $\trunkk{\ebpara}{S_x}$ in $\instance$ is the same as $\trunkk{\ebpara}{S'}$ in $\instance'$. Therefore, total disutility of $S_x$ in \instance is the same as total disutility of $S'$ in $\instance'$. Since $S_x \in \ruleof{\ebr}{}$, $S' \cap \ruleof{\ebr}{'} \neq \emptyset$.
\end{proof}

The next axiom, discount monotonicity \cite{talmon2019framework}, ensures that a winning project continues to win if it becomes less expensive.

\begin{definition}[Discount Monotonicity]\label{def: discountm}
A PB rule \br satisfies discount monotonicity if for any instance \instance and $x \in \winners{\br}{}$ such that $\cof{x} \geq 2$, we have $x \in \winners{\br}{'}$ whenever $\instance'$ is obtained from $\instance$ by reducing the cost of $x$ by 1.
\end{definition}
\begin{proposition}\label{the: discountm}
All the following rules satisfy discount monotonicity:
\begin{enumerate}
    \itemsep0em
    \item Any cost-worthy rule \cwr with $f(S)$ being $|S|$ or $\bool(|S|>0)$
    \item Any need-based rule \ebr where $\ebpara \in (0,1]$
\end{enumerate}
\end{proposition}
\begin{proof}
For an instance \instance and PB rule \br, let $x$ be a project in \winners{\br}{} such that $\cof{x} \geq 2$. Let $S_x$ be a set such that $S_x \in \ruleof{\br}{}$ and $x \in S_x$. Let $\instance'$ be an instance obtained from \instance by reducing \cof{x} by $1$. It is enough to prove that $S_x \in \ruleof{\br}{'}$.

We prove for cost-worthy rules and defer the rest to Appendix. Observe that for any agent $i$, $A_i$ remains the same in both \instance and $\instance'$. If $f(S) = |S|$, \scoreof{x} remains the same in \instance and $\instance'$. The utility of every set continues to be the same in $\instance'$. The utilities of sets also remain unchanged when $f(S) = \bool(|S)>0)$. Since $S_x \in \ruleof{\cwr}{}$, $S_x \in \ruleof{\cwr}{'}$.
\end{proof}

\begin{theorem}\label{prop: discountm}
The following rules do not satisfy discount monotonicity:
\begin{enumerate}
    \itemsep0em
    \item Any greedy-truncation rule with $f(S)$ as $|S|$, $\cof{S}$, and $\bool(|S|>0)$
    \item Any cost-worthy rule \cwr with $f(S)$ as $\cof{S}$.
    \item  Any need-based rule \ebr such that $\ebpara \in (1,\bud]$
\end{enumerate}
\end{theorem}
\begin{proof}
We start by looking at  greedy-truncation rules. Let $f(S)$ be $|S|$ or $\bool(|S|>0)$. Consider an instance \instance with budget $\bud > 3$, projects $\proj = \curly{a_1,a_2,a_3,a_4,a_5}$ costing $\curly{2,\bud-1,2,1,\bud}$ respectively, and four agents with preferences $a_1 \succ_1 a_2 \succ_1 others$, $a_2 \succ_2 a_3 \succ_2 others$, $a_4 \succ_3 a_5 \succ_3 others$, and $a_4 \succ_4 a_5 \succ_4 others$ respectively. Now, $\ruleof{\gtr}{} = \curly{\curly{a_1,a_4},\curly{a_2,a_4}}$ and $a_1 \in \winners{\gtr}{}$. If $\cof{a_1}$ is reduced by $1$, $\ruleof{\gtr}{'} = \curly{\curly{a_2,a_4}}$ and $a_1 \notin \winners{\gtr}{'}$.

Now, say $f(S) = \cof{S}$. Consider an instance \instance with budget $\bud > 3$, projects $\proj = \curly{a_1,a_2}$ each costing \bud, and two agents with preferences $a_1 \succ_1 a_2$ and $a_2 \succ_2 a_1$ respectively. Then, $\ruleof{\gtr}{} = \curly{\curly{a_1},\curly{a_2}}$ and $a_1 \in \winners{\gtr}{}$. However, if $\cof{a_1}$ is reduced by $1$, $\ruleof{\gtr}{'} = \curly{\curly{a_2}}$ and $a_1 \notin \winners{\gtr}{'}$.

Now, consider the \cwr rule with $f(S) = \cof{S}$. Consider an instance \instance with budget $\bud = \cwparaof{1}$, projects $\proj = \curly{a_1,a_2,d_1,\ldots,d_{m-2}}$ such that $\cof{a_1} = \cof{a_2} = \cwparaof{1}$ and for any $i \in \curly{1,\ldots,m-2}$ $\cof{d_i} = \cwparaof{i+2}+1$. Let there be two agents with preferences $a_1 \succ_1 a_2 \succ_1 d_1 \succ_1 \ldots \succ_1 d_{m-2}$ and $a_2 \succ_2 a_1 \succ_1 d_1 \succ_1 \ldots \succ_1 d_{m-2}$ respectively. Then, $\winners{\cwr}{} = \curly{a_1,a_2}$. However, if $\cof{a_1}$ is reduced by $1$, $\ruleof{\cwr}{'} = \curly{\curly{a_2}}$ and $a_1 \notin \winners{\cwr}{'}$.

Consider an instance \instance with budget $\bud$, projects $\proj = \curly{a_1,a_2}$ costing $\cof{a_1} = \max(2,\ceil{\ebpara})$ and $\cof{a_2} = \bud$. Say all agents have the same preference $a_1 \succ a_2$. Clearly, $\ruleof{\ebr}{} = \curly{\curly{a_1}}$. Now, decrease \cof{a_1} by $1$ to get $\instance'$. If $\ebpara > 2$, $\ruleof{\ebr}{'} = \curly{\curly{a_2}}$. Similarly when $1 < \ebpara \leq 2$, $\ruleof{\ebr}{'} = \curly{\curly{a_2}}$. Therefore, $a_1 \notin \winners{\ebr}{'}$. This completes the proof.
\end{proof}

The next axiom, limit monotonicity \cite{talmon2019framework}, requires that any winning project will continue to win if the budget is increased.

\begin{definition}[Limit Monotonicity]\phantomsection\label{def: limitm}
A PB rule \br satisfies limit monotonicity if for any instance \instance such that no project costs exactly $\bud+1$, we have $x \in \winners{\br}{'}$ whenever $x \in \winners{\br}{}$ and $\instance'$ is obtained from $\instance$ by increasing the budget by 1.
\end{definition}
\begin{proposition}\label{prop: limitm}
Any cost-worthy rule \cwr such that (i) $f(S)$ is $|S|$ or $\bool(|S|>0)$ and $\cwparaof{1} < \cwparaof{m}+2$ or (ii) $f(S) = \cof{S}$ and $\cwparaof{1} \leq 1$ satisfies limit monotonicity.
\end{proposition}
\vspace*{-0.7\baselineskip}
\begin{proof}
(i) Without loss of generality, assume that the first few entries in the worth vector are $\cwparaof{m}+1$ and the remaining are \cwparaof{m}. Consider a cost-worthy rule \cwr. Say $f(S) = |S|$. Take an instance \instance with budget \bud, and a project $x \in \winners{\cwr}{}$. Construct $\instance'$ by increasing the budget to $\bud\!+\!1$. Since $\prof$ and costs are unchanged, the score of any $S$ is the same in both \instance and $\instance'$. Consider any set $S' \in \ruleof{\cwr}{'}$ and any set $S_x \in \ruleof{\cwr}{}$ such that $x \in S_x$. For the sake of contradiction, assume $S_x \notin \ruleof{\cwr}{'}$. So, there exists $a \in S'$ and $b \in S_x$ such that $\scoreof{a}\!>\!\scoreof{b}$. However, since $S_x\!\in\!\ruleof{\cwr}{}$, $(S_x \!\setminus \!\curly{b})\! \cup\! \curly{a}$ must not be feasible in \instance (else, it would have strictly more utility than $S_x$). Hence, $\cof{S_x}\!+\!\cof{a}\!-\!\cof{b} \!>\! \bud$. Since $\cof{S_x} \!\leq\! \bud$, this implies, $\cof{b} \!<\! \cof{a}$. The score of any project whose cost is at most \cwparaof{m} is $n$. The score of any project whose cost is greater than $\cwparaof{m}+1$ is $0$. The score of any project whose cost is exactly $\cwparaof{m}\!+\!1$ will belong to $[0,n]$. Since $\scoreof{a} \!>\! \scoreof{b}$, one of these holds: (i) $\cof{a} \leq \cwparaof{m}$ and $\cof{b} \geq \cwparaof{m}+1$ (ii) $\cof{a} = \cwparaof{m}+1$ and $\cof{b} \geq \cwparaof{m}+1$. That is, $\cof{a} \leq \cwparaof{m}+1$ and $\cof{b} \geq  \cwparaof{m}+1$. This contradicts $\cof{b} < \cof{a}$. Now, say $f(S) = \bool(S>0)$. If there is any project costing at most $\cwparaof{m}$, it covers all the agents and the claim follows. If the layer selects only projects costing $\cwparaof{m}+1$ into each $A_i$, increasing \bud does not change the outcome.

(ii) Finally, say $f(S) = \cof{S}$ and $\cwparaof{1} \leq 1$. The layer selects only unit cost projects into each $A_i$. Thus, increasing \bud does not change the outcome.
\end{proof}
\begin{theorem}\label{the: limitm}
The following rules do not satisfy limit monotonicity:
\begin{enumerate}
    \itemsep0em
    \item Any greedy-truncation rule with $f(S)$ as $|S|$, $\cof{S}$, and $\bool(|S|>0)$
    \item Any cost-worthy rule \cwr with $f(S)$ as $\cof{S}$ with $\cwparaof{1} > 1$
    \item Any cost-worthy rule \cwr with $f(S)$ as $|S|$ or $\bool(|S|>0)$ and $\cwparaof{1} \geq \cwparaof{m}+2$.
    \item Any need-based rule \ebr
\end{enumerate}
\end{theorem}
The above theorem is proved in the appendix using counter examples. Notably, none of the PB rules studied by Talmon and Faliszewski \cite{talmon2019framework} satisfy limit monotonicity, though the axiom is introduced by them and is emphasized to be a very reasonable requirement in real-life scenarios. Interestingly, our \Cref{prop: limitm} recognizes a family of ordinal PB rules satisfying this axiom. Next, we study inclusion maximality \cite{talmon2019framework}, which implies that whenever some budget is remaining, agents would wish to use it.

\begin{definition}[Inclusion Maximality]\label{def: inclusionmax}
A PB rule \br satisfies inclusion maximality if for any instance \instance and $S, S' \in \feasible$ such that $S \subset S'$ and $S \in \ruleof{\br}{}$, we have $S' \in \ruleof{\br}{}$.
\end{definition}
\begin{proposition}\label{prop: inclusionmax}
All the following rules satisfy inclusion maximality:
\begin{enumerate}
    \itemsep0em
    \item Any greedy-truncation rule or cost-worthy rule where $f(S)$ is $|S|$, $\cof{S}$, or $\bool(|S|>0)$
    \item Any need-based rule \ebr
\end{enumerate}
\end{proposition}
\begin{proof}
A function $f$ on sets is said to be super-set monotone if $B \subseteq B'$ implies $f(B') \geq f(B)$. Notably, all the three functions $|S|$, $\cof{S}$, and $\bool(|S|>0)$ are super-set monotone. The claim follows for the layered approval rules. For any agent \ii, $S \subset S'$, and need-based rule \ebr, $\trunkk{\ebpara}{S} \geq \trunkk{\ebpara}{S'}$. Hence, the total disutility of $S'$ is at most that of $S$ and the claim follows.
\end{proof}
\subsubsection{Pro-Affordability}\label{sec: proafford}
We introduce a new axiom pro-affordability, which ensures that the PB rule always prefers a project that is less expensive and ranked higher by all agents.

\begin{definition}[Pro-affordability]\label{def: proafford}
A PB rule \br satisfies pro-affordability if for any instance \instance, $x \in \winners{\br}{}$, and $x' \in \proj$ such that $\cof{x'} < \cof{x}$ and $x' \suci x$ for every agent \ii, we have $x' \in \winners{\br}{}$.
\end{definition}
\begin{theorem}\label{the: proafford}
All the following rules satisfy pro-affordability:
\begin{enumerate}
    \itemsep0em
    \item Any greedy-truncation rule or cost-worthy rule with $f(S)$ as $|S|$ or $\bool(|S|>0)$
    \item A cost-worthy rule \cwr with $f(S) = \cof{S}$ and $\cwparaof{1} \leq 1$
    \item Any need-based rule \ebr where $\ebpara \in (0,1]$
\end{enumerate}
\end{theorem}
\begin{proof}
For any rule \br, an instance \instance, and a project $x \in \winners{\br}{}$, let $x'$ be a project such that $\cof{x'} < \cof{x}$ and $x' \suci x$ for every agent \ii. Let $S_x$ be a set such that $S_x \in \ruleof{\br}{}$ and $x \in S_x$. Let $S' = (S_x \setminus \curly{x}) \cup \curly{x'}$. Note that $S' \in \feasible$ since $S_x \in \feasible$ and $\cof{x'} < \cof{x}$. It is enough to prove that $S' \in \ruleof{\br}{}$. First, let us look at layered approval rules. Let \br be some greedy-truncation rule or cost-worthy rule with $f(S)$ as $|S|$ or $\cof{S}$. For any arbitrary agent \ii, if $x \in A_i$, then $x' \in A_i$ since $x' \suci x$. The utility of the set $S'$ is at least that of $S_x$ since $\scoreof{x'} \geq \scoreof{x}$. Therefore, $S' \in \ruleof{\br}{}$. Now, say $f(S) = \bool(|S|>0)$. For any arbitrary agent \ii, if $S_x \cap A_i \neq \emptyset$, then $S' \cap A_i \neq \emptyset$ since $x' \suci x$. Thus, the utility of $S'$ is at least as much as that of $S_x$ and $S' \in \ruleof{\br}{}$. Now say \br is \cwr with $f(S) = \cof{S}$ and $\cwparaof{1} \leq 1$. Since only unit cost projects are approved by the layer, claim follows.

Now, consider any need-based rule \ebr such that $\ebpara \in (0,1]$. Take any agent $i$. Since $\ebpara \leq 1$, $\trunkk{\ebpara}{S}$ is $\min_{a \in S}{\rof{a}}$. If $\rof{x'} \!<\! \trunkk{\ebpara}{S_x} \!\leq\! \rof{x}$, then $\trunkk{\ebpara}{S'} < \trunkk{\ebpara}{S_x}$. Else, $\trunkk{\ebpara}{S'} = \trunkk{\ebpara}{S_x}$. Hence, disutility of $S'$ is at most that of $S_x$. Since $S_x \in \ruleof{\ebr}{}$, $S' \in \ruleof{\ebr}{}$.
\end{proof}
\begin{proposition}\label{prop: proafford}
The following rules do not satisfy pro-affordability:
\begin{enumerate}
    \itemsep0em
    \item A greedy-truncation rule \gtr and a cost-worthy rule \cwr with $\cwparaof{1} > 1$ if $f(S) = \cof{S}$
    \item A need-based rule \ebr where $\ebpara \in (1,\bud]$
\end{enumerate}
\end{proposition}
\vspace*{-0.5\baselineskip}
Proof of the above proposition is deferred to Appendix.
\vspace*{-0.5\baselineskip}
\section{Conclusion}\label{sec: conclusion}
We introduced two classes of indivisible PB rules under weak rankings. The first class, layered approval votes, uses a layer to extract important information from weak rankings and convert weak rankings to approval votes. The first family of rules in this class, greedy-truncation rules, take advantage of weak rankings to greedily capture multiple knapsack votes of the agent. These rules also serve as a natural extension of top-$k$-counting rules of multi-winner voting and subsume multi-winner rules (bloc, $\alpha_k$-CC, etc.) as special cases. The second family of layered approval rules, cost-worthy rules, are the first PB rules to capture desirability of economical projects. Notably, these rules satisfy almost all the axioms and are computationally efficient. \Cref{tab: resultsla} summarises all the results for this class of rules.

Our second class of rules, the need-based rules, are the first indivisible PB rules to explicitly take into account the need of agents. Note that meeting the need of an agent is often the key for ensuring fairness \cite{bogomolnaia2005collective,aziz2019fair,aziz2014generalization,aziz2018rank,airiau2019portioning}. \Cref{tab: resultsnb} summarises all the results for this class of rules. In closing, our paper highlights the trade-offs among computational complexity, practical desirability, and axiomatic compliance of the proposed rules. 

We believe our work is a step towards a sound framework for indivisible PB under weak ordinal preferences. A natural direction is to study restricted domains in pursuit of tractable rules. Directions like providing axiomatic characterizations of our rules and identifying sub-classes of more desirable rules are interesting. Learning the optimal parameters of our rules based on real-world data is a promising direction.

\bibliography{paper.bib}

\begin{thebibliography}{10}

\bibitem{airiau2019portioning}
St{\'e}phane Airiau, Haris Aziz, Ioannis Caragiannis, Justin Kruger,
  J{\'e}r{\^o}me Lang, and Dominik Peters.
\newblock Portioning using ordinal preferences: Fairness and efficiency.
\newblock In {\em IJCAI}, 2019.

\bibitem{avritzer2000public}
Leonardo Avritzer.
\newblock Public deliberation at the local level: participatory budgeting in
  brazil.
\newblock In {\em Paper delivered at the Experiments for Deliberative Democracy
  Conference. Wisconsin}, 2000.

\bibitem{aziz2019fair}
Haris Aziz, Anna Bogomolnaia, and Herv{\'e} Moulin.
\newblock Fair mixing: the case of dichotomous preferences.
\newblock In {\em ACM EC}, pages 753--781, 2019.

\bibitem{aziz2021proportionally}
Haris Aziz and Barton~E Lee.
\newblock Proportionally representative participatory budgeting with ordinal
  preferences.
\newblock In {\em AAAI}, volume~35, pages 5110--5118, 2021.

\bibitem{aziz2018proportionally}
Haris Aziz, Barton~E Lee, and Nimrod Talmon.
\newblock Proportionally representative participatory budgeting: Axioms and
  algorithms.
\newblock In {\em AAMAS}, pages 23--31, 2018.

\bibitem{aziz2018rank}
Haris Aziz, Pang Luo, and Christine Rizkallah.
\newblock Rank maximal equal contribution: A probabilistic social choice
  function.
\newblock In {\em AAAI}, 2018.

\bibitem{aziz2021participatory}
Haris Aziz and Nisarg Shah.
\newblock Participatory budgeting: Models and approaches.
\newblock In {\em Pathways Between Social Science and Computational Social
  Science}, pages 215--236. Springer, 2021.

\bibitem{aziz2014generalization}
Haris Aziz and Paul Stursberg.
\newblock A generalization of probabilistic serial to randomized social choice.
\newblock In {\em AAAI}, 2014.

\bibitem{benade2021preference}
Gerdus Benade, Swaprava Nath, Ariel~D Procaccia, and Nisarg Shah.
\newblock Preference elicitation for participatory budgeting.
\newblock {\em Management Science}, 67(5):2813--2827, 2021.

\bibitem{betzler2013computation}
Nadja Betzler, Arkadii Slinko, and Johannes Uhlmann.
\newblock On the computation of fully proportional representation.
\newblock {\em Journal of Artificial Intelligence Research}, 47:475--519, 2013.

\bibitem{bogomolnaia2005collective}
Anna Bogomolnaia, Herv{\'e} Moulin, and Richard Stong.
\newblock Collective choice under dichotomous preferences.
\newblock {\em Journal of Economic Theory}, 122(2):165--184, 2005.

\bibitem{brandt2016handbook}
Felix Brandt, Vincent Conitzer, Ulle Endriss, J{\'e}r{\^o}me Lang, and Ariel~D
  Procaccia.
\newblock {\em Handbook of computational social choice}.
\newblock Cambridge University Press, 2016.

\bibitem{cabannes2004participatory}
Yves Cabannes.
\newblock Participatory budgeting: a significant contribution to participatory
  democracy.
\newblock {\em Environment and urbanization}, 16(1):27--46, 2004.

\bibitem{duddy2015fair}
Conal Duddy.
\newblock Fair sharing under dichotomous preferences.
\newblock {\em Mathematical Social Sciences}, 73:1--5, 2015.

\bibitem{elkind2017properties}
Edith Elkind, Piotr Faliszewski, Piotr Skowron, and Arkadii Slinko.
\newblock Properties of multiwinner voting rules.
\newblock {\em Social Choice and Welfare}, 48(3):599--632, 2017.

\bibitem{fairstein2021proportional}
Roy Fairstein, Reshef Meir, and Kobi Gal.
\newblock Proportional participatory budgeting with substitute projects.
\newblock {\em arXiv preprint arXiv:2106.05360}, 2021.

\bibitem{faliszewski2018multiwinner}
Piotr Faliszewski, Piotr Skowron, Arkadii Slinko, and Nimrod Talmon.
\newblock Multiwinner analogues of the plurality rule: axiomatic and
  algorithmic perspectives.
\newblock {\em Social Choice and Welfare}, 51(3):513--550, 2018.

\bibitem{freeman2021truthful}
Rupert Freeman, David~M Pennock, Dominik Peters, and Jennifer~Wortman Vaughan.
\newblock Truthful aggregation of budget proposals.
\newblock {\em Journal of Economic Theory}, 193:105234, 2021.

\bibitem{garey1979computers}
Michael~R Garey and David~S Johnson.
\newblock {\em Computers and intractability}, volume 174.
\newblock San Francisco: freeman, 1979.

\bibitem{goel2015knapsack}
Ashish Goel, Anilesh~K Krishnaswamy, Sukolsak Sakshuwong, and Tanja Aitamurto.
\newblock Knapsack voting.
\newblock {\em Collective Intelligence}, 1, 2015.

\bibitem{goel2019knapsack}
Ashish Goel, Anilesh~K Krishnaswamy, Sukolsak Sakshuwong, and Tanja Aitamurto.
\newblock Knapsack voting for participatory budgeting.
\newblock {\em ACM TEAC}, 7(2):1--27, 2019.

\bibitem{jain2020participatory}
Pallavi Jain, Krzysztof Sornat, and Nimrod Talmon.
\newblock Participatory budgeting with project interactions.
\newblock In {\em IJCAI}, pages 386--392, 2020.

\bibitem{jain2020participatoryg}
Pallavi Jain, Krzysztof Sornat, Nimrod Talmon, and Meirav Zehavi.
\newblock Participatory budgeting with project groups.
\newblock {\em arXiv preprint arXiv:2012.05213}, 2020.

\bibitem{lackner2021consistent}
Martin Lackner and Piotr Skowron.
\newblock Consistent approval-based multi-winner rules.
\newblock {\em Journal of Economic Theory}, 192:105173, 2021.

\bibitem{peters2020proportional}
Dominik Peters, Grzegorz Pierczy{\'n}ski, and Piotr Skowron.
\newblock Proportional participatory budgeting with cardinal utilities.
\newblock {\em arXiv preprint arXiv:2008.13276}, 2020.

\bibitem{pierczynski2021proportional}
Grzegorz Pierczy{\'n}ski, Piotr Skowron, and Dominik Peters.
\newblock Proportional participatory budgeting with additive utilities.
\newblock {\em Advances in Neural Information Processing Systems}, 34, 2021.

\bibitem{rey2020designing}
Simon Rey, Ulle Endriss, and Ronald de~Haan.
\newblock Designing participatory budgeting mechanisms grounded in judgment
  aggregation.
\newblock In {\em Proceedings of the International Conference on Principles of
  Knowledge Representation and Reasoning}, volume~17, pages 692--702, 2020.

\bibitem{rocke2014framing}
Anja R{\"o}cke.
\newblock {\em Framing citizen participation: participatory budgeting in
  France, Germany and the United Kingdom}.
\newblock Springer, 2014.

\bibitem{shapiro2017participatory}
Ehud Shapiro and Nimrod Talmon.
\newblock A participatory democratic budgeting algorithm.
\newblock {\em arXiv preprint arXiv:1709.05839}, 2017.

\bibitem{skowron2019axiomatic}
Piotr Skowron, Piotr Faliszewski, and Arkadii Slinko.
\newblock Axiomatic characterization of committee scoring rules.
\newblock {\em Journal of Economic Theory}, 180:244--273, 2019.

\bibitem{talmon2019framework}
Nimrod Talmon and Piotr Faliszewski.
\newblock A framework for approval-based budgeting methods.
\newblock In {\em AAAI}, volume~33, pages 2181--2188, 2019.

\end{thebibliography}
\bibliographystyle{plain}
\pagebreak
\section*{Appendix}\label{sec: app}
\newtheorem*{thediscountm_rep}{Theorem \ref{the: discountm}}
\begin{thediscountm_rep}
All the following rules satisfy discount monotonicity:
\begin{enumerate}
    \itemsep0em
    \item Any cost-worthy rule \cwr with $f(S)$ being $|S|$ or $\bool(|S|>0)$
    \item Any need-based rule \ebr where $\ebpara \in (0,1]$
\end{enumerate}
\end{thediscountm_rep}
\begin{proof}
    We prove the second point. Suppose \br is a need-based rule \ebr with $\ebpara \in (0,1]$. For any set $S$ and agent \ii, \trunkk{\ebpara}{S} is the rank of the best ranked project of $S$ in $\suci$. Consider an arbitrary agent \ii. If some project of $S_x$ is ranked before $x$ in $\suci$, \trunkk{\ebpara}{S_x} is the same in \instance and $\instance'$. If $x$ is the best ranked project of $S_x$ in $\suci$ in \instance, then \trunkk{\ebpara}{S_x} remains the same in $\instance'$ since $\cof{x}$ in $\instance'$ is at least $1$ and $\ebpara \leq 1$. Therefore, total disutility of $S_x$ in $\instance'$ is same as that in \instance. For any arbitrary agent \ii and a set $S$ without $x$ in it, \trunkk{\ebpara}{S} remains the same in $\instance'$. Hence, the total disutility of any set without $x$ remains the same in $\instance'$. Therefore, any set in $\ruleof{\ebr}{}$ with $x$ in it continues to be in $\ruleof{\ebr}{'}$ and hence $x \in \winners{\ebr}{'}$.
\end{proof}
\newtheorem*{thelimitm_rep}{\Cref{the: limitm}}
\begin{thelimitm_rep}
The following rules do not satisfy limit monotonicity:
\begin{enumerate}
    \item Any greedy-truncation rule with $f(S)$ as $|S|$, $\cof{S}$, and $\bool(|S|>0)$
    \item Any cost-worthy rule \cwr with $f(S)$ as $\cof{S}$ with $\cwparaof{1} > 1$
    \item Any cost-worthy rule \cwr with $f(S)$ as $|S|$ or $\bool(|S|>0)$ and $\cwparaof{1} \geq \cwparaof{m}+2$.
    \item Any need-based rule \ebr
\end{enumerate}
\end{thelimitm_rep}
\begin{proof}
\begin{enumerate}
    \item Consider a greedy-truncation rule \gtr with $f(S) = |S|$. Consider an instance \instance with budget $\bud > 3$, projects $\curly{a_1,a_2,a_3,a_4,a_5}$ respectively costing $\curly{2,\bud-1,3,1,\bud}$, and four agents with rankings: $a_1 \succ a_2 \succ \ldots \succ a_5$, $a_2 \succ a_3 \succ a_1 \succ a_4 \succ a_5$, $a_4 \succ a_5 \succ a_3 \succ a_1 \succ a_2$, and $a_4 \succ a_5 \succ a_3 \succ a_2 \succ a_1$. Then, $\ruleof{\gtr}{} = \curly{\curly{a_1,a_4},\curly{a_2,a_4}}$. Now if the budget is increased by $1$ to get $\instance'$, we have, $\ruleof{\gtr}{'} = \curly{\curly{a_2,a_4}}$. Hence, $a_1 \notin \winners{\gtr}{'}$. The same example also holds when $f(S) = \bool(|S|>0)$.

    Consider \gtr with $f(S) = \cof{S}$. Consider an instance \instance with budget $\bud > 4$, projects $\curly{a_1,a_2,a_3,a_4}$ respectively costing $\curly{2,\bud-1,2,\bud}$, and three agents with rankings: $a_1 \succ a_2 \succ a_3 \succ a_4$, $a_3 \succ a_4 \succ a_1 \succ a_2$, and $a_3 \succ a_4 \succ a_2 \succ a_1$. Then, $\ruleof{\gtr}{} = \curly{\curly{a_1,a_3}}$. Now if the budget is increased by $1$ to get $\instance'$, since $\bud+3 > 7$, we have, $\ruleof{\gtr}{'} = \curly{\curly{a_2,a_3}}$. Hence, $a_1 \notin \winners{\gtr}{'}$.
    \item Consider a cost-worthy rule \cwr with $\cwparaof{1} > 1$ and $f(S) = \cof{S}$. Consider an instance \instance with $\bud = 2\cwparaof{1}-2$. Say there are three projects $\curly{a_1,a_2,a_3}$ respectively costing $\curly{\cwparaof{1},\cwparaof{1}-1,1}$. Say there are $m-3$ projects $d_1,\ldots,d_{m-3}$ such that $\cof{d_i} = \cwparaof{i+1}+1$. Say there is only one agent and her ranking is $\curly{a_1,a_2,a_3} \succ d_1 \succ \ldots \succ d_{m-3}$. Then, $\ruleof{\cwr}{} = \curly{\curly{a_1,a_3}}$. If the budget is increased by $1$ to get $\instance'$, $\ruleof{\cwr}{'} = \curly{\curly{a_1,a_2}}$. Hence, $a_3 \notin \winners{\cwr}{'}$.
    \item Consider a cost-worthy rule \cwr with $\cwparaof{1} \geq \cwparaof{m}+2$ and $f(S) = |S|$ . Construct an instance \instance as follows: budget $\bud = \cwparaof{1}+\cwparaof{2}-1$, set of projects $\proj = \curly{a_1,\ldots,a_m}$, two agents have rankings $a_1 \succ \ldots \succ a_m$ and $a_1 \succ a_2 \succ a_m \succ a_4 \ldots \succ a_{m-1} \succ a_3$ respectively, and the costs are as follows:
    \begin{align*}
        \cof{a_i} &= \cwparaof{i}+1 \quad \forall i \in [3,m]\\
        \cof{a_1} &= \cwparaof{1}\\
        \cof{a_2} &= \cwparaof{2}
    \end{align*}
    Note that no project in $\proj$ costs exactly $\bud+1$. This is because, for any project $a_i$ such that $i \notin \curly{1,2}$, $\cof{a_i} = \cwparaof{i}+1$. Assume that this is equal to $\bud+1$, i.e., $\cwparaof{1}+\cwparaof{2}-\cwparaof{i}=1$. Since $\cwparaof{2} \geq \cwparaof{i} \geq 0$, this implies $\cwparaof{1} = 1$. This contradicts $\cwparaof{1} \geq \cwparaof{m}+2$. Also, note that $a_1$ and $a_2$ have a score of $2$ each while $a_m$ might have a score of $1$. All other projects have a score of $0$.
    
    Consider the set $\curly{a_2,a_m}$. We know that $\cof{\curly{a_2,a_m}} = \cwparaof{2}+\cwparaof{m}+1$. Since $\cwparaof{m} \leq \cwparaof{1}-2$, $\cof{\curly{a_2,a_m}} \leq \bud$. Since \curly{a_1,a_2} is infeasible, score of \curly{a_2,a_m} is optimal. Therefore, $\curly{a_2,a_m} \in \ruleof{\cwr}{}$ and $a_m \in \winners{\cwr}{}$. Now, construct an instance $\instance'$ by increasing \bud to $\cwparaof{1}+\cwparaof{2}$. Now, \curly{a_1,a_2} is feasible and has the optimal maximum score of $4$. Hence, $a_m \notin \winners{\cwr}{'}$.

    Now, consider \cwr with $f(S) = \bool(|S|>0)$. Say we have an instance \instance with $\bud = 2\cwparaof{1}-1$. Let $\curly{a_1,a_2,a_3}$ be projects respectively costing $\curly{\cwparaof{1}-1,\cwparaof{1},\cwparaof{1}}$. Let $\curly{d_1,\ldots,d_{m-3}}$ be projects such that $\cof{d_i} = \cwparaof{i+1}+1$. Let there be six agents such that the ranking of one agent is $a_1 \succ d_1 \succ \ldots \succ d_{m-3} \succ a_2 \succ a_3$, ranking of two agents is $a_2 \succ d_1 \succ \ldots \succ d_{m-3} \succ a_3 \succ a_1$, and ranking of remaining three agents is $a_3 \succ d_1 \succ \ldots \succ d_{m-3} \succ a_2 \succ a_1$. Clearly, $\ruleof{\cwr}{} = \curly{\curly{a_1,a_3}}$. If we increased budget by $1$ to get $\instance'$, then $\ruleof{\cwr}{'} = \curly{\curly{a_2,a_3}}$. Thus, $a_1 \notin \winners{\cwr}{'}$.
    \item Now, consider any need-based rule \ebr. First we prove that if $\ebpara$ is independent of \bud and is a constant, it is easy to construct an instance where \ebr does not satisfy limit monotonicity. Construct an instance with budget $\bud = 2\ceil{\ebpara}-1$, projects $\proj = \curly{a_1,a_2,a_3}$ costing $\ceil{\ebpara}$, $2\ceil{\ebpara}-1$, and $\ceil{\ebpara}$ respectively, and two agents with preferences $a_1 \succ_1 a_2 \succ_1 a_3$ and $a_3 \succ_2 a_2 \succ_2 a_1$ respectively. Note that no project in $\proj$ costs $\bud+1$. Clearly no non-singleton set is feasible and disutility of every singleton set is $4$. Therefore, $\ruleof{\ebr}{} = \curly{\curly{a_1},\curly{a_2},\curly{a_3}}$. Now, construct an instance $\instance'$ by increasing \bud to $2\ceil{\ebpara}$. Now though \curly{a_2} is feasible, it has a disutility of $4$ whereas the feasible set \curly{a_1,a_3} has an optimal disutility of $2$. Therefore, $\ruleof{\ebr}{'} = \curly{\curly{a_1,a_3}}$ and $a_2 \notin \winners{\ebr}{'}$.

    Now, we construct two instances respectively to prove that \ebr does not satisfy limit monotonicity when $\ceil{\ebpara}$ is in $(0,\frac{\bud+1}{2}]$ and $(\frac{\bud+1}{2},\bud]$. Let $W$ be an odd number greater than $3$. First, assume $\ceil{\ebpara} \leq \frac{\bud+1}{2}$. Construct an instance as follows: $\bud = W$ and $\proj = \curly{a_1,a_2,a_3}$ costing $\ceil{\ebpara}$, \bud, and $\bud-\ceil{\ebpara}+1$ respectively. Say two agents have the following preferences: $a_1 \succ_1 a_2 \succ_1 a_3$ and $a_3 \succ_2 a_2 \succ_2 a_1$. Since, $\ceil{\ebpara} \leq \frac{\bud+1}{2}$, $\bud - \ceil{\ebpara} + 1 \geq \ceil{\ebpara}$. Therefore, $\ruleof{\ebr}{} = \curly{\curly{a_1},\curly{a_2},\curly{a_3}}$ where each set has disutility $4$. Now, construct an instance $\instance'$ by increasing \bud to $\bud+1$. Then, $\ruleof{\ebr}{'} = \curly{\curly{a_1,a_3}}$ and $a_2 \notin \winners{\ebr}{'}$.
    
    Finally, say $\ceil{\ebpara} > \frac{\bud+1}{2}$. Now construct an instance with budget $\bud = W$, projects $\proj = \curly{a_1,a_2,a_3}$ costing $\ceil{\ebpara}-1$, $\bud-\ceil{\ebpara}+2$, and $\bud-\ceil{\ebpara}+1$ respectively. Without loss of generality, we can assume $\ceil{\ebpara} > 1$ (we already proved that if \ebpara is independent of \bud, we can construct an instance to prove the claim). Say all the agents have the preference $a_1 \succ a_2 \succ a_3$. Since \bud is odd and $\ceil{\ebpara} > \frac{\bud+1}{2}$, $\ceil{\ebpara} > \frac{\bud+2}{2}$. Therefore, $\bud-\ceil{\ebpara}+2 < \ceil{\ebpara}$. By definition, $\ebr$ and $\operatorname{D}_{\ceil{\ebpara}}$ are equivalent since all costs are positive integers. Any singleton set has a disutility of $4n$ for the rule $\operatorname{D}_{\ceil{\ebpara}}$. We can see that $\curly{a_1,a_2}$ is infeasible. Hence, $\curly{a_1,a_3} \in \ruleof{\ebr}{}$ and $a_3 \in \winners{\ebr}{}$. Now, construct an instance $\instance'$ by increasing \bud to $\bud+1$. Then, $\ruleof{\ebr}{'} = \curly{\curly{a_1,a_2}}$ and $a_3 \notin \winners{\ebr}{'}$. This completes the proof.
\end{enumerate}
\end{proof}

\newtheorem*{propproafford_rep}{\Cref{prop: proafford}}
\begin{propproafford_rep}
The following rules do not satisfy pro-affordability:
\begin{enumerate}
    \item A greedy-truncation rule \gtr and a cost-worthy rule \cwr with $\cwparaof{1} > 1$ if $f(S) = \cof{S}$
    \item A need-based rule \ebr where $\ebpara \in (1,\bud]$
\end{enumerate}
\end{propproafford_rep}
\begin{proof}
\begin{enumerate}
    \item Say $f(S) = \cof{S}$. Consider the greedy-truncation rule \gtr. Consider an instance \instance with budget $\bud > 2$, four projects $\curly{a_1,a_2,a_3,a_4}$ respectively costing $\curly{1,\bud-1,1,\bud}$. Say there are three agents such that one agent has a ranking $a_1 \succ a_2 \succ a_3 \succ a_4$ and two agents have a ranking $a_3 \succ a_4 \succ a_1 \succ a_2$. Clearly $\ruleof{\gtr}{} = \curly{\curly{a_2,a_3}}$ though $a_1$ is preferred over $a_2$ by all the agents and $\cof{a_1} < \cof{a_2}$.

    Now consider the cost-worthy rule \cwr with $\cwparaof{1} > 1$. Consider an instance with budget $\bud = \cwparaof{1}$. Say there are two projects $a_1$ and $a_2$ respectively costing $\cwparaof{1}-1$ and $\cwparaof{1}$. Also, there are $m-1$ projects such that $\cof{d_i} = \cwparaof{i+1}+1$. Say there is only agent and her preference is $\curly{a_1,a_2} \succ d_1 \succ \ldots \succ d_{m-1}$. Then, $\ruleof{\cwr}{} = \curly{\curly{a_2}}$ though $\cof{a_1} < \cof{a_2}$.
    \item Take the \ebr rule where $\ebpara > 1$. Since $\ebpara>1$, disutility of \curly{a_1} is $4n$ whereas disutility of \curly{a_2} and \curly{a_3} are $2n$ and $3n$ respectively. Therefore, $\ruleof{\ebr}{} = \curly{\curly{a_2}}$ and $a_1 \notin \winners{\ebr}{}$. This completes the proof.
\end{enumerate}
\end{proof}
\end{document}